\documentclass[prb, twocolumn, nofootinbib]{revtex4-2}

\usepackage[unicode=true,colorlinks=true]{hyperref}

\usepackage{amsthm,amsmath,amsfonts,amssymb}%
\usepackage{bbold}
\usepackage{bm}
\usepackage{graphicx}
\usepackage[dvipsnames]{xcolor}
\usepackage{color}

\newcommand{\KP}{$\bm k\cdot\bm p$}
\newcommand{\rmi}{{\rm i}}
\newcommand{\rme}{{\rm e}}

\newcommand{\up}{\uparrow}
\newcommand{\down}{\downarrow}
\newcommand{\ket}[1]{\left|{#1}\right\rangle}

\newcommand{\brakt}[2]{\left\langle{#1}\middle\vert{#2}\right\rangle}

\DeclareMathOperator{\diag}{diag}
\newcommand{\ebasis}{\bm{\mathcal{E}}}

\newcommand{\gf}{\texorpdfstring{$g$}{g}-factor}
\newcommand{\gfs}{\texorpdfstring{$g$}{g}-factors}

\begin{document}

\title{Theory for electron and hole fine structure and \\ Land\'e g-factors in lead chalcogenide nanowires}
\author{I.D.~Avdeev}
\email{ivan.avdeev@mail.ioffe.ru}
\author{M.O.~Nestoklon}
\affiliation{Ioffe Institute, 194021 St. Petersburg, Russia}

\begin{abstract}
Using the atomistic tight-binding method in combination with symmetry analysis and extended effective mass theory we derive a phenomenological model for the fine structure of the ground electron and hole states in $[111]$-grown PbX, X=S,Se hexagonal and cylindrical nanowires.
Projection of the electron (hole) states calculated in empirical tight-binding method to the basis of valley states enables the explicit parametrization of the valley mixing Hamiltonian, which is essential to obtain correct combinations of the valley states for electron (hole) ground levels analytically. 
The effective Hamiltonian of valley mixing allows to convert the intravalley components of the \gfs\ tensors to the Zeeman splittings of electron and hole levels in nanowires.
Tensors of the intravalley electron (hole) Land\'e \gfs, parameters of the valley mixing Hamiltonian as well as the valley and anisotropic splitting energies and velocity matrix elements are also calculated.
\end{abstract}

\maketitle

\section{Introduction}

Lead chalcogenide (PbS, PbSe, PbTe) nanostructures are known for excellent tunability of their operating wavelength in the infrared and near-infrared spectral range which makes them ideal platform for infrared optoelectronics and energy harvesting devices \cite{Sukhovatkin09,Oh2015,Iacovo2016}.
In particular, semiconductor nanowires (NWs) are considered to be an ideal planform for next generation high-performance electronics \cite{Jia2019}.
Recently, it was also demonstrated that lead chalcogenide NWs have variety of potential applications ranging from a new type of field effect transistor to Josephson junctions for quantum computers \cite{Kim2017,Schellingerhout2022,Li2024}.
Large static dielectric constants \cite{Zemel1965} and Land\'e \gfs\ \cite{Rabii1968,Leolup1973,Avdeev2023g} of lead chalcogenides open the way to tune the physical properties of NWs both by varying the gate voltage and external magnetic field while keeping the NWs tolerant to charged defects and impurities.

The major challenge for accurate theoretical description of PbX (X=S, Se, Te) NWs is the complex multivalley band structure of lead chalcogenides.
Both conduction and valence band extrema in PbX are located at the four inequivalent anisotropic $L$ valleys \cite{Svane10} yielding the intricate $8$-fold degenerate electron and hole levels.
This degeneracy is partially lifted in the NWs due to the anisotropy of the effective masses \cite{Bartnik10} and the intervalley scattering at the surface \cite{Avdeev2017}.
Even though there are several theoretical works on PbX NWs, including the empirical tight-binding (ETB) calculations \cite{Paul11,Avdeev2017,Avdeev2019}, \KP\ theory \cite{Bartnik10,Cao2022} and density functional theory (DFT) approach \cite{Aryal2021}, the accurate parametrization of the intervalley scattering is missing, in contracst to cubic PbX quantum dots \cite{Avdeev2020}.
In the NWs the intervalley scattering is possible between all $L$ valleys projected onto equivalent points of the Brillouine zone of the NW and its magnitude strongly depends on the NW shape and growth direction \cite{Avdeev2019}.

Here we focus on hexagonal and cylindrical PbX NWs grown along the $[111]$ crystallographic axis as they provide more flexibility to control the fine strucuture of electron and hole levels via interplay of the anisotropic and valley splittings of states.
Using the technique proposed in Ref.~\cite{Avdeev2024JETP} we untangle the valley structure of states and obtain the electron (hole) states localized in certain $L$ valleys.
This allows us to obtain explicit paramterization of the valley mixing Hamiltonian and calculate the intravalley parameters such as effective masses, confinement energies, anisotropic splittings, velocity matrix elements and Land\'e \gfs\ directly in the framework of the ETB method.
We also put forward a simplified formalism for the anisotropic \KP\ theory and propose a phenomenological model for the electron and hole dispersion in the $[111]$ PbX NWs in the vicinity of the band extrema.

\section{Structure of hexagonal nanowires}
\label{sec:NW}

\begin{figure}
  \includegraphics[width=0.5\linewidth]{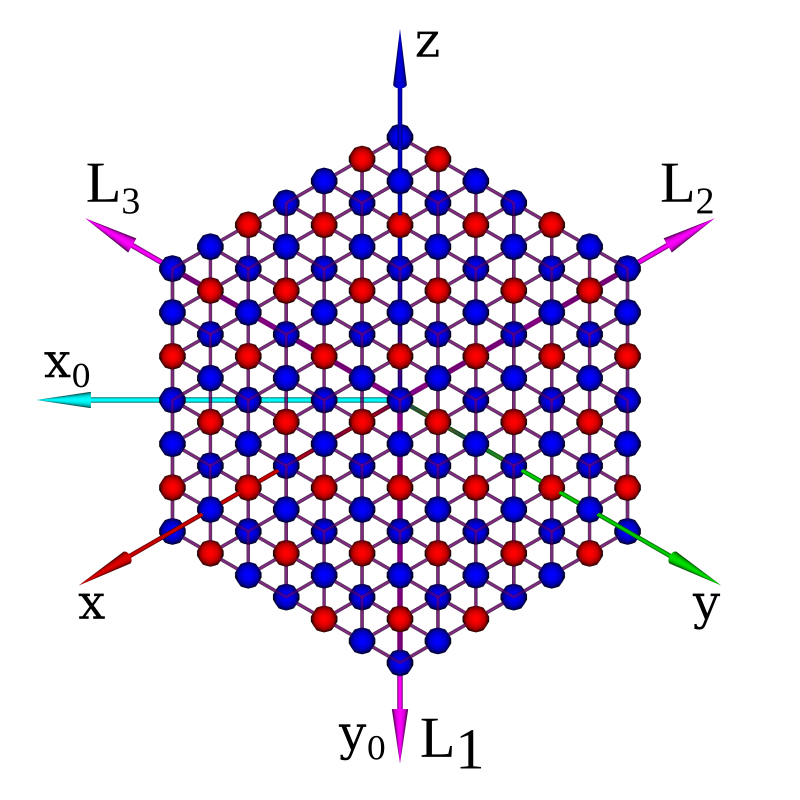}
  \caption{Atomic structure of the hexagonal PbX nanowire with the size parameter $N=6$ projected onto the $(111)$ cross section plane. Cations (Pb) are shown by red balls, anions (S,Se) by blue ones. The crystallographic axes $x,y,z$ are shown by red, blue and green arrows. The $x_0$ and $y_0$ axes of the NW coordinates frame are shown by blue and cyan arrows. Projections of the inclined valleys onto the $(111)$ cross section plane are shown by cyan arrows.}
  \label{fig:NW_N6}
\end{figure}

Compared to cylindrical PbX NWs grown along the $[111]$ axis \cite{Avdeev2017} the hexagonal NWs have much simpler atomic structure with surface terminated by neutral $\{110\}$ faces.
The $(111)$ cross section of a hexagonal PbX NW and its orientation with respect to the crystallographic axes and $L$ valleys are shown in Fig~\ref{fig:NW_N6}.

In the calculations, elementray cell of the NW is cut from ideal bulk PbX crystal with abrupt surface without additional passivation or relaxation.
Same approach was used in Refs.~\cite{Avdeev2017} and \cite{Avdeev2019}.
Specifically, in Ref.~\cite{Avdeev2019} it was shown that these distortions of the NW surface have almost no effect on the electron (hole) fine structure of the NWs.
The NW axis $[111]$ passes through the cation (Pb atom) located at the origin of the crystallographic coordinates, as a result the NWs are centrosymmetric.
In Fig~\ref{fig:NW_N6} the crystallographic axes $x\parallel[100]$, $y\parallel[010]$ and $z\parallel[001]$ are shown by red, blue and green arrows, respectively.
The NW-related coordinates frame is chosen as
\begin{equation}
  \label{eq:cf0}
  x_0 \parallel [1\bar10],\quad
  y_0 \parallel [11\bar2],\quad
  z_0 \parallel [111].
\end{equation}
The axes $x_0$ and $y_0$ are shown by cyan and magenta arrows in Fig.~\ref{fig:NW_N6}.
The axis $z_0$ (not shown in the plot) points towards the reader.
This coordinates frame is used throughout the paper if not stated otherwise explicitlty.

The size of the hexagon is defined by the integer number $N$ (size parameter).
It describes the number of atomic layers surrounding the cental axis and sets the distance $l_{x_0}=N a_0/\sqrt2$ between the opposing hexagon edges along the $x_0$ axis.
In Fig.~\ref{fig:NW_N6} a NW with the size parameter $N=6$ is shown.
We also introduce effective diameter of NWs,
\begin{equation}
  D_{\text{eff}}(N) = \frac{3^{\frac14}}{\sqrt{\pi}}Na_0,
\end{equation}
as the diameter of the circle with the same area as the $(111)$ hexagonal NW cross section.
For $N=6$ the effective diameter $D_{\text{eff}}\approx2.85$ nm for PbS ($a_0=5.9$ \AA) and $2.95$ nm for PbSe ($a_0=6.1$ \AA).
The lattice constants are taken from Ref.~\cite{Poddubny2012}.

Translational symmetry of the hexagonal $[111]$ PbX NWs is given by the vector
\begin{equation}\label{eq:AB}
  \bm A = a_0(1,1,1),\quad
\end{equation}
see Ref.~\cite{Avdeev2017} for details.
The elementary cell of the $[111]$ PbX NWs has layered structure and consists of six $(111)$ alternating cation and anion atomic layers.

The size of the one-dimensional Brillouin zone of the $[111]$-grown PbX NWs equals to the length of the reciprocal lattice vector $B=|\bm B|$, 
\begin{equation}\label{eq:AB_B}
  \bm B = \frac{2\pi}{a_0\sqrt3} \frac{1}{\sqrt3}(1,1,1),
\end{equation}
The band extrema are located at projections of the $L$ valleys onto the NW axis \cite{Avdeev2017,Avdeev2019}.
In case of the $[111]$-grown PbX NWs all four $L$ valleys project to equivalent points, therefore all valley states can mix due to the intervalley scattering \cite{Avdeev2017}.
In this work we choose the $L$ valleys as
\begin{equation}\label{eq:LL}
  L_0 \parallel [111],\quad
  L_1 \parallel [\bar1\bar11],\quad
  L_2 \parallel [1\bar1\bar1],\quad
  L_3 \parallel [\bar11\bar1].
\end{equation}
The $L_0$ axis is aligned with the NW axis, the three others $L_{i}, i=1,2,3$ are inclined.
Inclined valleys are shown in Fig.~\ref{fig:NW_N6} by magenta arrows.
Projection of the $L_1$ valley onto the $(111)$ plane is parallel to the $y_0$ axis.
The wave vectors of the $L_{\mu}$ valleys are $\bm k_{\mu}=\sqrt3\frac{\pi}{a_0}\bm n_{\mu}$, where $\bm n_{\mu}$ are the unit vectors in Eq.~\eqref{eq:LL}.
Different angles between the $[111]$ NW axis and longitudinal, $L_0$, and inclined, $L_i, i=1,2,3$, valleys result in additional valley splitting due to the mass anisotropy of the $L$ valleys. This effect can be captured within the framework of the \KP\ theory~\cite{Bartnik10}.

The point symmetry of the hexagonal NWs is $D_{3d}$.
This group has three generators: $a=C_{3z_0}$ is the $2\pi/3$ rotation around the NW axis $z_0$, $b=C_{2x_0}$ is the $2\pi/2$ rotation around the axis $x_0$ and inversion $C_i$.
The origin of the point symmetry operations is chosen at the cation at the origin of the crystallographic coordinates frame.
This choice guarantees the odd (even) parity of conduction (valence) band states \cite{Avdeev2017}.

\section{Tight-binding}

\subsection{Method}

In the empirical tight-binding (ETB) method, the $i$-th electron wave function $\Phi_i (\bm{r})$ is expanded in the basis of orthogonal atomic-like functions $\phi_{\sigma}(\bm{r})$ \cite{Lowdin50}:
\begin{equation}\label{eq:TB_psi_r}
  \Phi_i (\bm{r}) = \sum_{n \sigma} C_{n \sigma}^i \phi_{\sigma}(\bm{r} - \bm{r}_{n} )\:,
\end{equation}
where $n$ enumerates the atoms and the index $\sigma$ runs through different orbitals.
We use the $sp^3d^5s^*$ variant of the method \cite{Jancu98} with twenty orbitals per atom and parameters taken from \cite{Avdeev2017}.
In this basis, the Schr\"odinger equation reduces to the eigenvalue problem for a sparse matrix:
\begin{equation}\label{eq:TB_Ham_r}
  \sum_{n',\varsigma} H_{n\sigma,n'\varsigma} C^i_{n'\varsigma} = \epsilon_i C^i_{n\sigma} \,.
\end{equation}
Here $\epsilon_i$ is the energy of $i$-th electron state.
Magnetic field is added following Ref.~\cite{Graf95} by (i) adding diagonal ($n=n'$, intraatomic) Zeeman splittings of the spin-up and spin-down orbitals proportional to the magnetic field and (ii) by multiplying the off-diagonal ($n\ne n'$, interatomic) matrix elements by the dot product of the vector potential $\bm A$ and the vector of the corresponding chemical bond $\bm r_n-\bm r_{n'}$. The choice of vector potential is discussed below. 
The band-edge solutions of~\eqref{eq:TB_Ham_r} are found using the Thick-Restart Lanczos algorithm \cite{Wu00}.

\subsection{Valley states}

In this section we modify the valley untangling procedure proposed in Ref.~\cite{Avdeev2024JETP} to apply it for the $[111]$-PbX nanowires with $D_{3d}$ symmetry.

\begin{figure}
  \includegraphics[width=.8\linewidth]{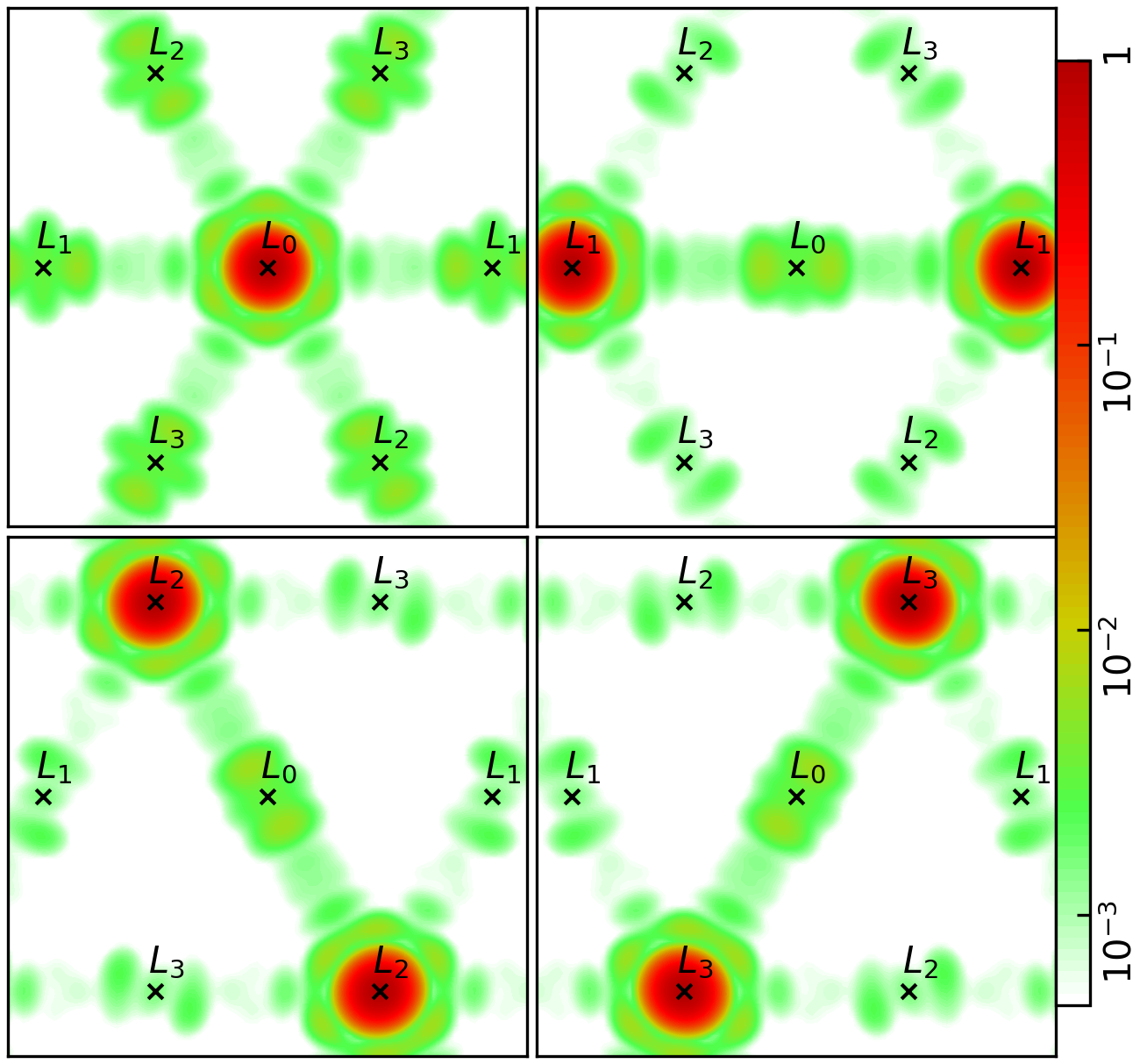}
  \caption{Spin averaged local density of electron valley states in PbS NW with $N=6$ (see Fig.~\ref{fig:NW_N6}) in reciprocal space (kDOS) in the $(111)$ plane passing through the $L_0$ valley. Projections of the $L$ valleys onto this plane are idicated by ``$\times$''. The amplitude of the density is shown by color in logarithmic scale.}
  \label{fig:kDOS_2D}
\end{figure}

As the input for the valley untangling procedure we use the Fourier images of the electron and hole states obtained via diagonalization of the tight-binding Hamiltonian \eqref{eq:TB_Ham_r}.
To adopt the procedure for nanowires with lower than cubic $D_{3d}$ symmetry we relate the states in inclined $L_2$ and $L_3$ valleys via $C_{3z}$ rotation of the states in $L_1$ valley, see Appendix~\ref{app:symm}.
Then the procedure from Ref.~\cite{Avdeev2024JETP} is applied straightforwardly to obtain independent sets of states in $L_0$ and $L_1$ valleys as linear combinations of the ETB NW eigenstates:
\begin{equation}\label{eq:TB2VB}
  \ebasis_{\text{TB}}^b W^b =\ebasis^b\,,\quad b=c,v\,,
\end{equation}
where $\ebasis_{\text{TB}}^b$ is rectangular matrix with its columns being the tight-binding eigenstates corresponding to the ground electron ($b=c$) or hole ($b=v$) levels $\left\{\ebasis_{\text{TB}}^b\right\}_{i,n\sigma}=C_{n\sigma}^{i=c_1..c_8}$ and $W^b$ is the $8\times 8$ matrix obtained via the valley untangling procedure.
The resulted states
\begin{equation}\label{eq:eb}
  \ebasis^b=(\ebasis_0^b,\ebasis_1^b,\ebasis_2^b,\ebasis_3^b),\quad b=c,v,
\end{equation}
are the pairs of electron and hole spin-like states each localized near one of the $L$ valleys.
To illustrate this in Fig.~\ref{fig:kDOS_2D} we show the continuous local density in reciprocal space (kDOS) \cite{Nestoklon16_Ge,Avdeev2017} of electron states in the same hexagonal PbS NW with $N=6$ ($D_{\text{eff}}\approx 2.85$ nm) as shown in Fig.~\ref{fig:NW_N6}.
The kDOS is computed following Ref.~\cite{Avdeev2017} in the $(111)$ plane passing through the $L_0$ valley.
The kDOS does not depend on the (pseudo)spin index of the valley states.
In Fig.~\ref{fig:kDOS_2D} projections of the $L$ valleys (or equivalent points differ by a bulk reciprocal lattice vector) are indicated by ``$\times$'' symbols with corresponding label.

In Fig.~\ref{fig:kDOS_2D} one can clearly see the sharp peak of the kDOS near exactly one (the kDOS is periodic with the period given by bulk reciprocal lattice vectors) specific projection of the $L$ valley onto the $(111)$ plane.
This indicates the successful numerical optimization of the unknown rotation angles between repetitive irreducible subspaces comprising the ground electron/hole level fine structure in the hexagonal PbS NW.
The logarithmic scale is chosen for the plot to depict the small admixture of high energy levels (with higher angular momentum) in other valleys.
These admixtures of the kDOS are responsible for the second in valley mixing order corrections to the intervalley terms in effective Hamiltonians.
They are out of the scope of the present paper.

In the valley untangling procedure \cite{Avdeev2024JETP} the kDOS, a quadratic characteristic of the electron (hole) states, is optimized.
As a result the relative signs of electron and hole states and the relative signs of longitudinal and inclined valley states can not be unambigously extracted.
There is also additional symmetry allowed degree of freedom related to the orientation of the valley pseudospinors in inclined valleys along or against the NW axis, see Appendix~\ref{app:symm} for details.
The relative sign between conduction and valence band states can be adjusted by fixing the sign of the interband velocity matrix elements.
The relative sigh between longitudinal and inclined valleys can be adjusted by fixing the sign of the corresponding matrix element of the valley mixing Hamiltonian.
The orientation of the inclined valleys pseudospinors can be adjusted by tracing corresponding components of the intravalley tensors of \gfs.

\section{Results}

Our main result is the explicit parametrization of the effective Hamiltonian describing the fine structure of the ground electron (hole) confinement level in the vicinity of the band extremum in the hexagonal $[111]$-grown PbX NWs in the presence of external magnetic field $\bm B$.
This Hamiltonian reads as
\begin{equation}
  \label{eq:Hvc}
  \hat H(\bm B, k_z) = 
  \begin{pmatrix}
    H^c(\bm B, k_z) & \hbar \hat V_z k_z \\
    \hbar \hat V_z^{\dag} k_z & H^v(\bm B, k_z)
  \end{pmatrix}
\end{equation}
where $k_z$ is the wave number along the NW axis $[111]$, $k_z=0$ is the band extremum corresponding to the projection of the $L_0$ valley to the NW axis, $\hat V_z$ is the $z$ component of interband velocity operator and $\hat H^{b}(\bm B, k_z)$, $b=c(v)$ is the electron (hole) part of the Hamiltonian
\begin{equation}
  \label{eq:Hb}
  \hat H^{b}(\bm B, k_z) = \hat H_{\text{aniso}}^{b} + \hat H_{\text{VM}}^{b} + \hat H^b(\bm B)+\hat H^b(k_z).
\end{equation}
Here $\hat H_{\text{aniso}}^{b}$ describes both the quantum confinement and anisotropic splitting of the valley states, $\hat H_{\text{VM}}^{b}$ describes the valley mixing of states, $\hat H^b(\bm B)$ describes the interaction with magnetic field (we consider only linear in $\bm{B}$ terms) and $\hat H^b(k_z) \propto k_z^2$.
The last term describes correction to the electron (hole) effective masses along the NW axis due to the contributions from remote bands.

\subsection{Anisotropic splitting}
\label{sec:an}

\begin{figure}
  \includegraphics[width=1\linewidth]{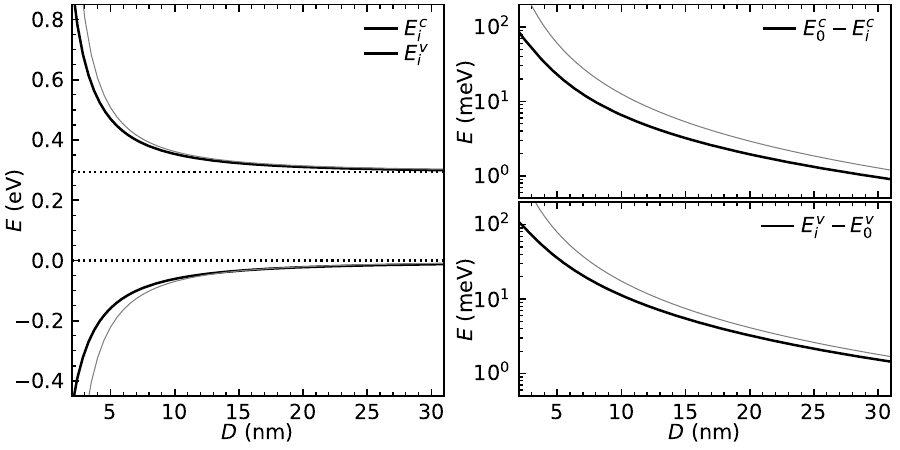}
  \caption{Ground confinement energies $E_i^{c(v)}$ and anisotropic splittings $\pm(E_0^{c(v)}-E_i^{c(v)})$ in conduction and valence band in hexagonal PbS NWs as a function of the NW diameter. Results of tight-binding calculations are shown by thick black lines, calculations in the framework of anisotropic \KP\ theory are shown by thin grey lines. Dotted lines on the left plot show valence band maximum (chosen as zero) and the low-temperature conduction band minimum $E_{\text{CBM}}=0.294$ eV of the bulk PbS crystal.}
  \label{fig:EA_PbS_N}
\end{figure}

In the basis of valley pseudospinors \eqref{eq:eb} the anisotropic part of the Hamiltonian takes the form
\begin{equation}
  \label{eq:Han}
  \hat H_{\text{aniso}}^{b} = \diag(E_0^b,E_i^b,E_i^b,E_i^b) \otimes \mathbb 1.
\end{equation}
Here $E_0$ is the confinement energy in longitudinal valley and $E_i$ is the confinement energy of inclined valleys, $\mathbb 1$ is the $2\times2$ unit matrix.
Due to the smaller effective masses in $(111)$ plane the confinement energy in longitudinal valley is larger than the one in inclined valleys, therefore $|E_0^b|>|E_i^b|, b=v,c$ \cite{Bartnik10}.
It allows us to refer $E_i^{c(v)}$ as the ground confinement energy and $\pm (E_0^{c(v)}-E_i^{c(v)})>0$ as the anisotropic splitting in conduction (valence) band.
These energies can be easily calculated in the framework of \KP\ theory \cite{Bartnik10}.
In tight-binding these energies are defined as
\begin{equation}\label{eq:Emub}
  E_{\mu}^b = \frac{\sum_{s=1}^8 \rho_s^b(\bm k_{\mu}) \epsilon_s^b}{\sum_{s=1}^8 \rho_s^b(\bm k_{\mu})},
  \quad \mu=0,1,2,3,
  \quad b=c,v.
\end{equation}
Here $s$ enumerates the first eight conduction (valence) band tight-binding eigenstates, $\epsilon_s^{c(v)}$ are their energies and $\rho_s^{c(v)}(\bm k_{\mu})$ are the values of their local density at the four $L_{\mu}$ valleys.
Eq.~\eqref{eq:Emub} traces contributions to the valley energy $E_{\mu}^{c(v)}$ among all states of the valley multiplet. Therefore
Eq.~\eqref{eq:Emub} allows to rule out the effect of the valley mixing to obtain the confinement energy and anisotropic splitting.

Results of the calculation are shown in Fig.~\ref{fig:EA_PbS_N}.
The ground confinement energies, $E_i^{c(v)}$, are shown in the left part of Fig.~\ref{fig:EA_PbS_N}, anisotropic splittings are shown on the right in logarithmic scale.
Results of tight-binding calculations are shown by black lines, calculations in the framework of anisotropic \KP\ theory are shown by thin grey lines.

\subsection{Valley mixing}
\label{sec:vm}

\begin{figure}
  \includegraphics[width=1\linewidth]{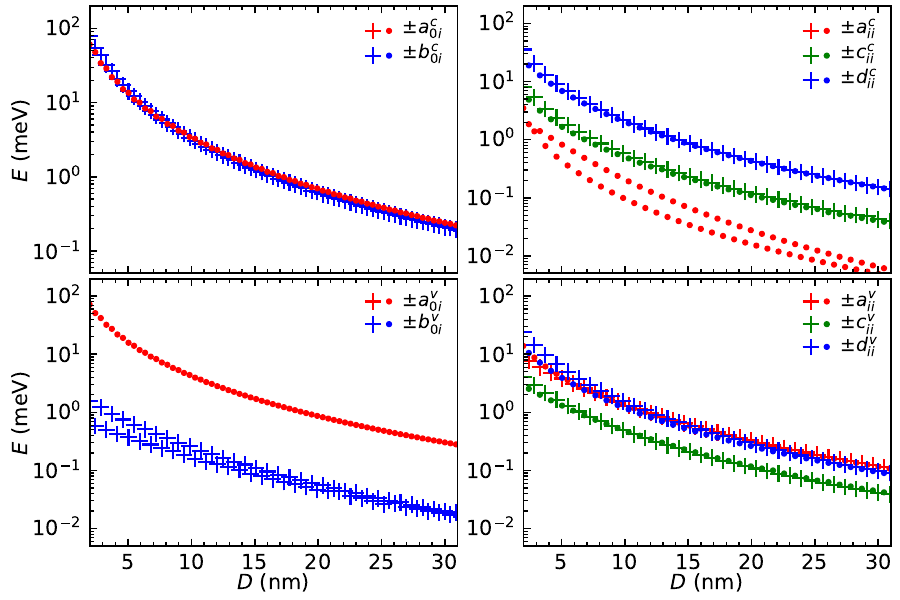}
  \caption{Constants of the valley mixing Hamiltonian in conduction (top) and valence (bottom) bands. Left panels show the constants describing the valley mixing between longitudinal and inclined valleys, right ones show the intervalley scattering constants between inclined valleys. Signs of the constants are indicated by symbols: ``$+$'' for positive and ``$\bullet$'' for negative values.}
  \label{fig:VS_PbS_N}
\end{figure}

The second term in Eq.~\eqref{eq:Hb} is the Hamiltonian of the valley mixing $\hat H_{\text{VM}}^b$.
This Hamiltonian mixes different valley states therefore in the basis of valley pseudospinors, Eq.~\eqref{eq:eb}, its diagonal (intravalley) blocks are zeros $H_{\text{VM}}^{\mu\mu}=\ebasis_{\mu}^{\dag}\hat H_{\text{VM}}\ebasis_{\mu}=\mathbb 0$.
Since the longitudinal $\mu=0$ and inclined valleys $\mu=1,2,3$ are independent with respect to the NW point symmetry $D_{3d}$, there are only two independent $2\times2$ blocks: $H_{\text{VM}}^{01}$ and $H_{\text{VM}}^{12}$.
The first block is related to $H_{\text{VM}}^{02}$ and $H_{\text{VM}}^{03}$ via
\begin{equation}
  H_{\text{VM}}^{02} = D_{\frac12}(C_{3z}) H_{\text{VM}}^{01},\quad
  H_{\text{VM}}^{03} = D_{\frac12}^2(C_{3z}) H_{\text{VM}}^{01}
\end{equation}
due to the choice of the bases in $L_2$ and $L_3$ valleys, see Eq.~\eqref{eq:E23E1} in Appendix~\ref{app:symm}.
One may obtain similar expressions for the inclined valleys $H_{\text{VM}}^{\mu\nu}$, $\mu,\nu=1,2,3$
\begin{equation}
  H_{\text{VM}}^{12} = H_{\text{VM}}^{23} = -H_{\text{VM}}^{31}
\end{equation}
and using the $C_{2x}$ rotation
\begin{equation}\label{eq:Hvm13symm}
  H_{\text{VM}}^{13} = - \sigma_x H_{\text{VM}}^{12} \sigma_x.
\end{equation}
The other blocks are simply conjugated $H_{\text{VM}}^{\mu\nu}={H_{\text{VM}}^{\nu\mu}}^{\dag}$.
Our tight-binding calculations suggest that $H_{\text{VM}}^{01}$ may be parametrized by two real constants
\begin{equation}
  \label{eq:Hvm01}
  H_{\text{VM}}^{01} = a_{0i}\mathbb 1 + \rmi b_{0i}\sigma_x
\end{equation}
describing ``spin conserving'' and ``spin-flip'' intervalley scattering.
The mixing between inclined valleys $H_{\text{VM}}^{12}$ is characterized by three real constants
\begin{equation}
  \label{eq:Hvm12}
  H_{\text{VM}}^{12} = a_{ii}\mathbb 1 + \rmi (c_{ii}\sigma_y + d_{ii} \sigma_z).
\end{equation}
Here $\sigma_x,\sigma_y$ and $\sigma_z$ are the Pauli matrices, $\mathbb 1$ is the $2\times2$ unit matrix.

To compute the valley mixing Hamiltonian we use the following transformation
\begin{equation}
  \label{eq:Hvm}
  \hat H_{\text{VM}}^b = W^{b\dag}\diag(\epsilon_{b_1},\epsilon_{b_2},\ldots,\epsilon_{b_8})W^b - \hat H_{\text{aniso}}^b.
\end{equation}
Here $\epsilon_{b_i}, i=1\ldots8$ are the first eight tight-binding eigenenergies in conduction ($b=c$) or valence ($b=v$) band, $\hat H_{\text{aniso}}^{c(v)}$ is the anisotropic part of the Hamiltonian \eqref{eq:Han} and $W^{c(v)}$ defined in \eqref{eq:TB2VB}.
Both Eqs.~\eqref{eq:Han} and \eqref{eq:Hvm} are computed at the band extremum $k_z=0$.

Results of the calculations, the constants \eqref{eq:Hvm01} and \eqref{eq:Hvm12}, are shown in Fig.~\ref{fig:VS_PbS_N}.
To better illustrate the scaling of the constants they are plotted on logarithmic scale. 
The sign of the valley mixing constants is represented by symbols: negative values are shown by ``$\bullet$'', positive ones by ``$+$''.
One can clearly see that in conduction band the normal and spin-flip scattering amplitudes between $L_0$ and $L_1$ valleys are almost identical, while in other cases the spin-flip amplitudes are significantly smaller.
Same holds for cylindrical PbS and both type of PbSe NWs, but oscillations of the valley mixing constants in cylindrical NWs are much more pronounced.

\subsection{Linear in \texorpdfstring{$k$}{k} splittings}
\label{sec:k1}
\begin{figure}
  \includegraphics[width=1\linewidth]{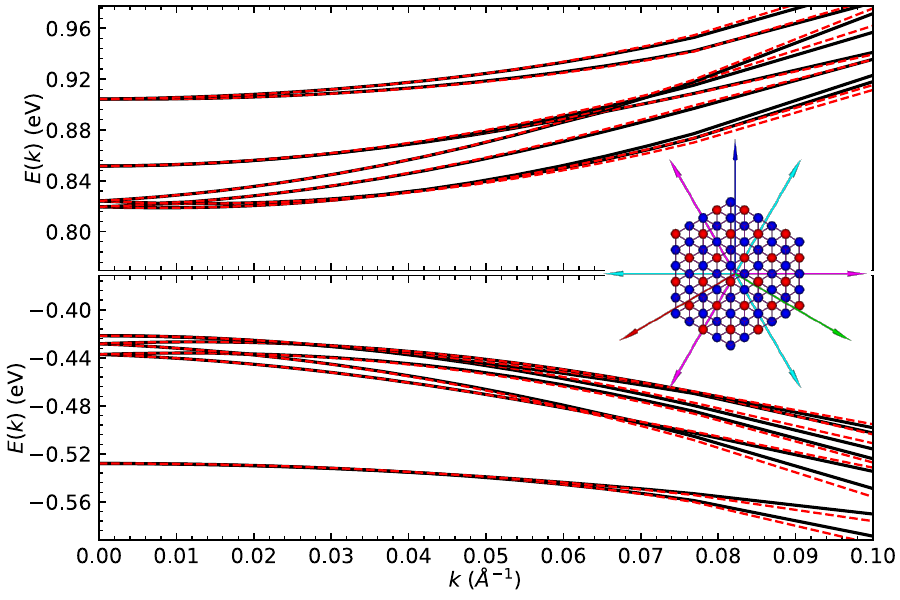}
  \caption{An example dispersion of the hexagonal PbS nanowire without inversion. The nanowire cell and all auxiliary axes are shown in the miniature.
  Parameters of the linear in $k$ terms in the valley mixing Hamiltonian, Eqs.~\eqref{eq:Hvm01k} and \eqref{eq:Hvm12k}, are: 
  $\tilde a_{0i}^v=-79.1$, 
  $\tilde b_{0i}^v=180.0$, 
  $\tilde a_{ii}^v=88.5$, 
  $\tilde c_{ii}^v=101.7$, 
  $\tilde d_{ii}^v=62.7$, 
  $\tilde a_{0i}^c=-89.1$, 
  $\tilde b_{0i}^c=61.3$, 
  $\tilde a_{ii}^c=152.0$, 
  $\tilde c_{ii}^c=52.8$, 
  $\tilde d_{ii}^c=41.2$
  (meV$\cdot$\AA).
  Solid black lines show dispersion calculated in tight-binding.
  The red dashed lines show dispersion calculated in the phenomenological model with the linear in $k$ terms in the valley mixing Hamiltonian.
  }
  \label{fig:HVMk}
\end{figure}

When the nanowire lacks the center of inversion, additional linear in $k$ terms appear in the NW dispersion. 
In Ref.~\cite{Avdeev2017} it has been argued that the magnitude of these splittings is proportional to the value of the valley mixing of the longitudinal and inclined valley states, in a close analogy with the spin splittings in SiGe quantum wells \cite{Nestoklon2006,Nestoklon2008}.
In our case of the $[111]$ PbX NWs these linear terms can be parametrized by five additional real constants.
These constants enter the blocks~\eqref{eq:Hvm01} and \eqref{eq:Hvm12} of the valley mixing Hamiltonian as follows:
\begin{subequations}
\begin{equation}
  \label{eq:Hvm01k}
  H_{\text{VM}}^{01}\to H_{\text{VM}}^{01}+ 
  k \left(
    \tilde a_{0i}\sigma_z + \tilde b_{0i}\sigma_y
  \right),
\end{equation}
\begin{equation}
  \label{eq:Hvm12k}
  H_{\text{VM}}^{12}\to H_{\text{VM}}^{12}+
  k\left(
    \tilde a_{ii}\sigma_z + \tilde c_{ii} \sigma_y+ \rmi d_{ii} \mathbb 1
  \right).
\end{equation}
\end{subequations}
An example dispersion of the $[111]$ PbS NW without center of inversion is shown in Fig.~\ref{fig:HVMk}.
The ETB calculated dispersion shown by solid black lines agrees well with our extended \KP\ model calculations shown by red dashed lines.
The $(111)$ cross section of the NW cell is shown in the miniature. 
Additional parameters of the linear in $k$ splittings are given in the Figure caption.
The rest of the \KP\ parameters are discussed below.

\subsection{Tensors of intravalley \gfs}
\label{sec:gfs}

To calculate the tensors of $g$-factors in NWs we use following gauge for the vector potential in the NW coordinates frame \eqref{eq:cf0}:
\begin{equation}\label{eq:A}
  \bm A = \left(-\frac{B_{z_0}y_0}2,\frac{B_{z_0}x_0}2, B_{x_0}y_0-B_{y_0}x_0\right)_{x_0y_0z_0}.
\end{equation}
This gauge ensures the proper translation symmetry of the tight-binding Hamiltonian in the external magnetic field \cite{Graf95} and allows to avoid integration over $z$ axis in the framework of \KP\ theory.

In ETB, we calculate the matrix elements of $\hat H(\bm B)$ in small magnetic field ($B\le$0.1 T) in the basis of tight-binding eigenstates at the band extremum at zero magnetic field $\hat H(k_z=0, \bm B=0)$.
Next, we compute only linear in $\bm{B}$ part of the Hamiltonian
\begin{equation}\label{eq:HBraw}
  \delta \hat H(\bm B) = \hat H(k_z=0,\bm B\ne0)-\mathrm{diag}\left\{ \epsilon_i \right\}.
\end{equation}
Here we subtract the energies $\epsilon_i$ in the absence of magnetic field from the matrix $\hat H(k_z=0,\bm B\ne0)$ of the tight-binding Hamiltonian.
The matrix $\hat H(k_z=0,\bm B\ne0)$ is computed in the basis of the zero field eigenstates.

Using the $W^{b\dag}\delta \hat H^b(\bm B)W^b$ transformation \eqref{eq:TB2VB} for each band $b=c,v$ we obtain the intravalley linear in $\bm B$ blocks of the Hamiltonian
\begin{equation}\label{eq:g_definition}
  \delta H^{b,\mu\mu}(\bm B) = 
  {\ebasis_{\mu}^{b}}^{\dag}\delta \hat H^b(\bm B){\ebasis_{\mu}^{b}}
  = 
  \frac{\mu_B}2 g_{\alpha\beta}^{\mu,b} \sigma_{\alpha} B_{\beta}.
\end{equation}
There are two independent tensors of \gfs\ in each band $b=c,v$: $g_{\alpha\beta}^{0,b}$ and $g_{\alpha\beta}^{i,b}\equiv g_{\alpha\beta}^{1,b}$, which correspond to the longitudinal valley $L_0$ and inclined valley $L_1$.
Interaction with magnetic field in inclined valleys $L_2$ and $L_3$ is given by $\delta H^{b,11}(C_{3z_0}^{-1}\bm B)$ and $\delta H^{b,11}(C_{3z_0}^{-2}\bm B)$, respectively.

\begin{figure}
  \includegraphics[width=1\linewidth]{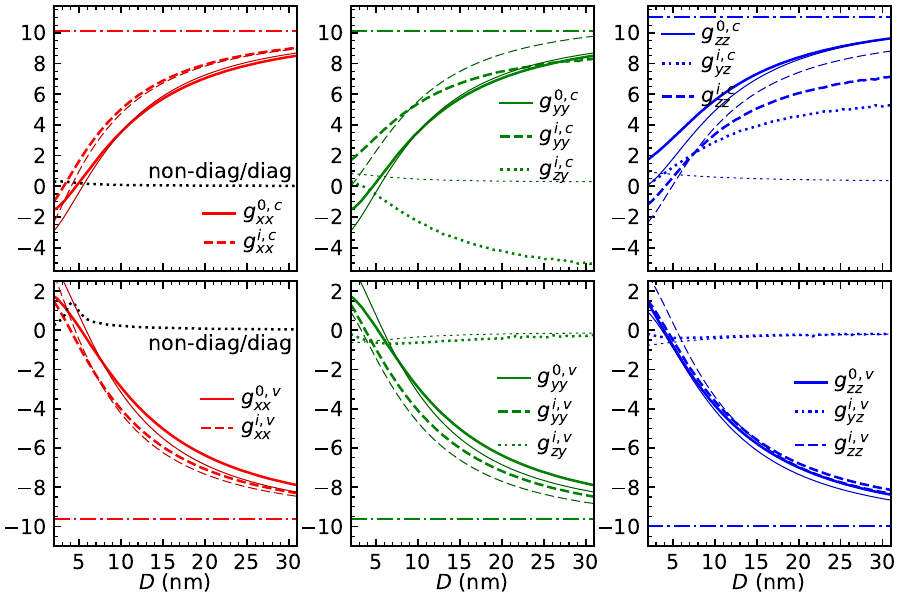}
  \caption{Components of the intravalley conduction and valence band Land\'e \gfs\ tensors in hexagonal PbS NWs as a function of the NW diameter. Results of tight-binding calculations are shown by thick solid, dashed and dotted lines, calculations in the framework of the anisotropic \KP\ theory are shown by thin lines with same style. Dashdot lines in each plot show the corresponding value of intravalley \gf\ in bulk PbS. Black dotted lines on the left show conduction (top left panel) and valence (bottom left panel) band ratio of intervalley to intravalley components of the \gfs\ tensors.}
  \label{fig:HB_PbS_N}
\end{figure}

All nonzero components of the \gfs\ tensors in conduction and valence band in hexagonal PbS NWs are shown in Fig.~\ref{fig:HB_PbS_N}.
Results of the tight-binding calculations are shown by thick solid, dashed and dotted lines.
Calculations within the framework of anisotropic \KP\ theory, see Appendix \ref{app:kp}, are shown by thin lines with the same linestyles.
Horizontal dashdot lines show the components of the bulk tensor of \gfs.
Results of \KP\ and tight-binding calculations agree well with each other, convergence to the bulk values is clear.

The intervalley interaction with magnetic field ${\ebasis_{\mu}^{c(v)}}^{\dag}\delta \hat H(\bm B){\ebasis_{\nu\neq\mu}^{c(v)}}$ can be neglected as its contribution to the Zeeman splittings of the electron (hole) levels is negligibly small given the overall accuracy of the model (see the next section).

\subsection{Zeeman splitting of electrons and holes}

In this section we use the results of Sec.~\ref{sec:an}, Sec.~\ref{sec:vm} to compute the Zeeman splittings of electron (hole) levels in hexagonal PbS nanowires using the i) tight-binding and ii) \KP\ values for the intravalley \gfs\ tensors discussed the previous section.

In the $[111]$-PbX NWs with $D_{3d}$ symmetry the electron (hole) states at the band extremum split into four doublets forming $3\Gamma_4\oplus\Gamma_{56}$ reducible representation of the point group, see Appendix.~\ref{app:symm}.
The energies of these doublets depend on anisotropic splitting \eqref{eq:Han} and the five constants of the valley mixing Hamiltonian \eqref{eq:Hvm}.
Therefore in these NWs the order of the doublets energy levels and their exact composition of the valley states are not easy to predict.
However, the $D_{3d}$ symmetry allows for description of the Zeeman splittings by phenomenological diagonal tensors of \gfs
\begin{equation}\label{eq:g456}
  \hat g_{\Gamma_4} = \diag(g_t,g_t,g_l)
  ,\quad 
  \hat g_{\Gamma_{56}} = \diag(0,0,\tilde g_l)
\end{equation}
for the $\Gamma_4$ and $\Gamma_{5,6}$ doublets, respectively.
These are the most important \gfs\ as they can be observed in experiments.
Below we do not trace the valley structure of repetitive $\Gamma_4$ states and only follow their energies in ascending (descending) order in conduction (valence) band. 
\begin{figure}
  \includegraphics[width=0.25\linewidth]{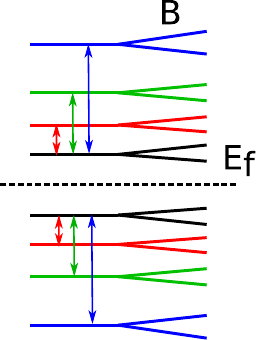}
  \caption{Scheme of anisotropic, valley and Zeeman splitting of electron and hole doublets in PbX NWs. The Zeeman splittings are $\Delta E_{t(l)} = \mu_B g_{t(l)} B$. Dashed line indicate the Fermi level. 
  }
  \label{fig:split_scheme}
\end{figure}

\begin{figure}[t]
  \includegraphics[width=1\linewidth]{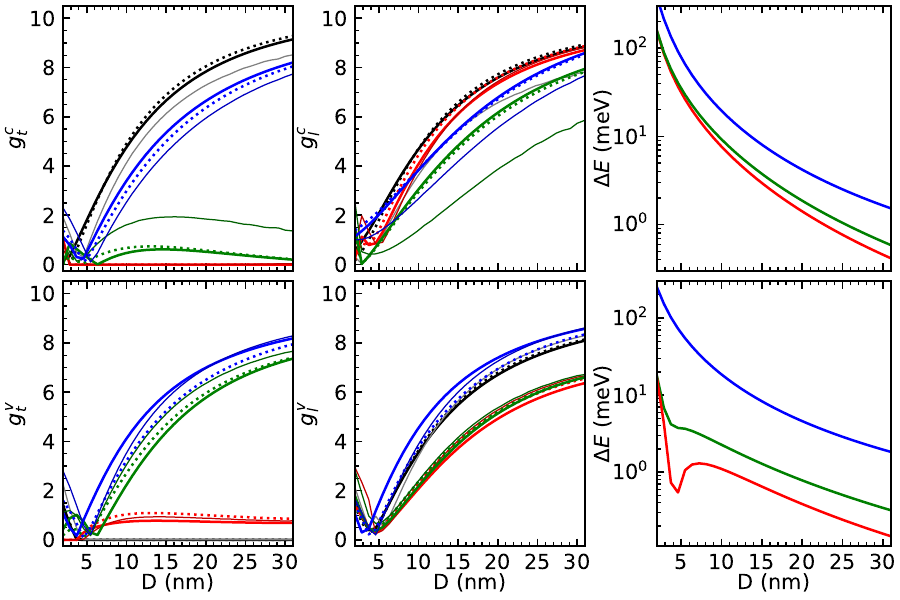}
  \caption{Absolute values of effective \gfs\ of electron (top) and hole (bottom) doublets and energy splittings betwee doublets (right) in hexagonal PbS NWs with even $N$ as a function of NW diameter. The energy splittings are indicated by color with respect to the scheme~\ref{fig:split_scheme}.
  Results of tight-binding calculations with diagonal and non-diagonal by the valley index terms in $H(\bm B)$ are shown by thick solid lines and without non-diagonal terms by thick dotted lines. Calculations with $H(\bm B)$ obtained from the anisotropic \KP\ model are shown by thin solid lines.
  }
  \label{fig:gg_HB_even}
\end{figure}

\begin{figure}[t]
  \includegraphics[width=1\linewidth]{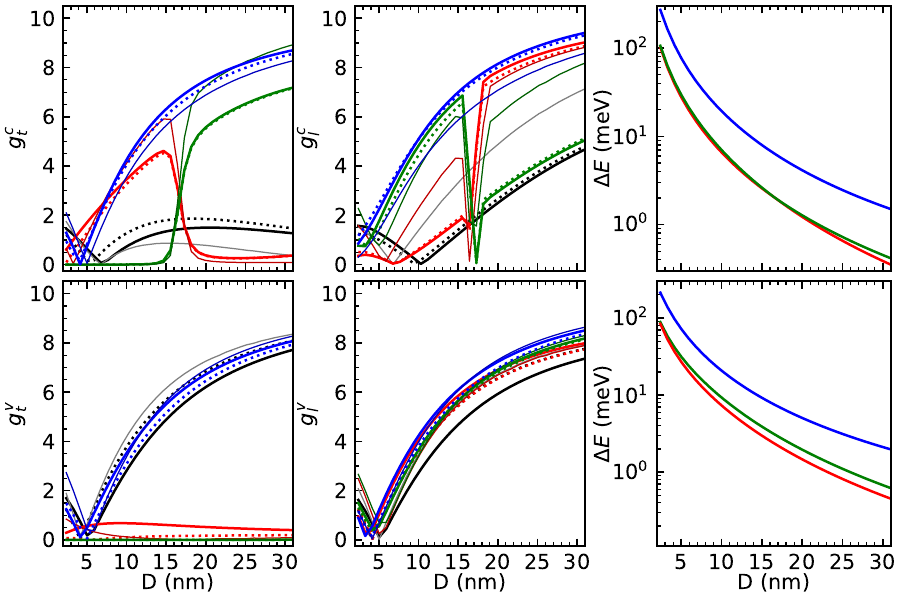}
  \caption{Same as in Fig.~\ref{fig:gg_HB_even}, but for odd $N$. The feature of conduction band \gfs\ at $D\approx17$ nm arise due to the intersection of the second (red) and third (green) electron doublets.}
  \label{fig:gg_HB_odd}
\end{figure}

Results of the calculations are shown in Figs.~\ref{fig:gg_HB_even} and~\ref{fig:gg_HB_odd} for hexagonal PbS NWs with even and odd values of size parameter $N$, respectively.
Such a division is chosen to avoid overloading of the plots since there are distinct odd-even oscillations of the valley mixing constants in hexagonal NWs, see Fig.~\ref{fig:VS_PbS_N}.
Apart from even-odd oscillations, the valley mixing constants are smooth functions of the NW diameter.
As a result, the values of the effective \gfs\ \eqref{eq:g456} also do not oscillate.
The absolute values of the effective \gfs\ \eqref{eq:g456} in conduction and valence bands are calculated in ETB as the ratios of the doublet splitting energies and the magnitude of the (small) magnetic field
\begin{equation}\label{eq:g_ETB}
  g_{l(t)} = \frac{|\Delta \epsilon_{l(t)}^i|}{\mu_B B_{l(t)}}.
\end{equation}
Here $\Delta \epsilon_{l(t)}^i$ is the splitting of the $i$'th tight-binding energy doublet when magnetic field is applied along (perpendicular) to the NW axis.
The energy levels $i$ are encoded with colors as shown in scheme \ref{fig:split_scheme}.
In the right panels of Figs.~\ref{fig:gg_HB_even} and \ref{fig:gg_HB_odd} we show the total energy splitting between the lowest (highest) and $i$'th electron (hole) doublet without magnetic field, see scheme~\ref{fig:split_scheme}.
One can clearly see the feature in Fig.~\ref{fig:gg_HB_odd} in PbS NWs with odd $N$ number in conduction band \gfs\ at $D\approx17$ nm.
At this point the second (red) and third (green) electron levels intersect.
In cylindrical NWs the oscillations of the valley splittings are much more chaotic, see supplemental information in Appendix~\ref{app:SI}.

In Figs.~\ref{fig:gg_HB_even} and \ref{fig:gg_HB_odd} thick solid lines show results of the full tight-binding calculations \eqref{eq:g_ETB}.
Dotted lines show the tight-binding results when only the valley diagonal terms in $H(\bm B)$ are taken into account.
One can see that solid and dotted lines differ marginally which proves that the non-diagonal by the valley index terms in $H(\bm B)$ can be omitted.
Thin solid lines show the same values but with intravalley components of \gfs\ tensors taken from the anisotropic \KP\ model.

\subsection{Matrix elements of velocity}
\label{sec:VME}

\begin{figure}
  \includegraphics[width=1\linewidth]{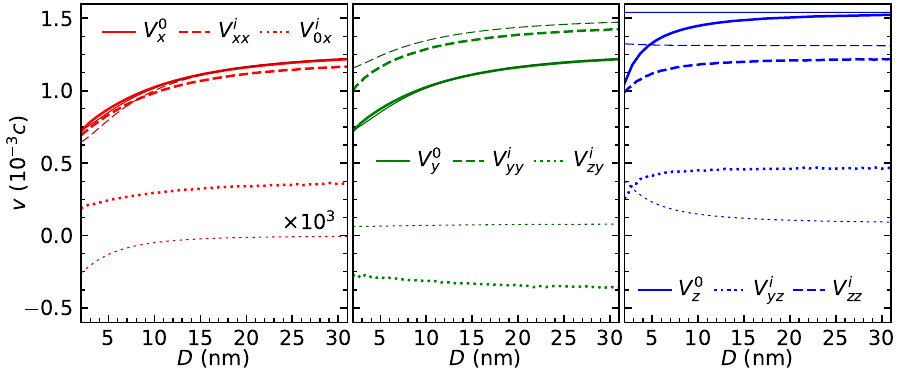}
  \caption{Interband intravalley velocity matrix elements in hexagonal PbS NWs as a function of the NW diameter. Results of tight-binding calculations are shown by thick solid, dashed and dotted lines, \KP\ data are shown by thin lines with same line styles. The $V_{0x}^i$ component (thin dotted line in the left panel) is magnified by $\times10^3$ to make it visually appear in the plot.}
  \label{fig:VME_PbS_N}
\end{figure}

Longitudinal velocity matrix elements along the NW axis $z$ enter the non-diagonal blocks of the effective Hamiltonian \eqref{eq:Hvc}.
We also compute all the other components of the velocity matrix elements in PbX NWs.

In ETB method the velocity operator is represented as the commutator
\begin{equation}
  \hat{\bm{V}} = \frac{i}{\hbar} \left[ H, \bm{r} \right]
\end{equation}
of the tight-binding Hamiltonian $H$, Eq.~\eqref{eq:TB_Ham_r}, and the coordinate operator in diagonal approximation $\{\bm{r}\}_{n\sigma,n'\sigma'}= \bm{r}_{n} \delta_{nn'}\delta{\sigma\sigma'}$, see Refs.~\cite{RamMohan93,Ivchenko06,Avdeev2020SiGe,Avdeev2020} for details.

First we consider the longitudinal valley $L_0$ aligned with the NW axis.
Since the electron and hole states in $L_0$ valley are spin-like, then the interband intravalley velocity matrix elements have the following form
\footnote{In the NW coordinates frame \eqref{eq:cf0}.}
\begin{equation}
  \label{eq:VME0}
  {\ebasis_0^{c}}^{\dag} \hat{\bm V} \ebasis_0^v = 
  \left(V_t^0\sigma_x,V_t^0\sigma_y,V_l^0\sigma_z\right),
\end{equation}
where $V_t^0$ and $V_l^0$ are the transverse and longitudinal components of the velocity.
The interband intravalley velocity matrix elements in inclined valleys, $L_i, i=1,2,3$ have more complicated form due to the misalignment of the valley's and the NW axes.
Tight-binding calculations show that the velocity matrix elements in inclined valley $L_1$\footnote{In $L_2$ and $L_3$ valleys the matrix elements are the same if coordinates frame \eqref{eq:cf0} is rotated by $C_{3z}$ and $C_{3z}^2$, respectively.} can be parametrized as
\begin{multline}
  \label{eq:VMEi}
  {\ebasis_i^{c}}^{\dag} \hat{\bm V} \ebasis_i^v = \\
  \left(\rmi V_{0x}^i\mathbb 1+V_{xx}^i\sigma_x,
  V_{yy}^i\sigma_y+V_{zy}^i\sigma_z,
  V_{zz}^i\sigma_z+V_{yz}^i\sigma_y\right).
\end{multline}
Results of the tight-binding calculations for hexagonal PbS NWs are shown in Fig.~\ref{fig:VME_PbS_N} by thick color lines in the thousandth of speed of light units ($10^{-3}$ c).
In Fig.~\ref{fig:VME_PbS_N} results of the calculation in the framework of the anisotropic \KP\ theory are show by thin lines with same line styles.
Results of the calculations suggest that anisotropic \KP\ theory is capable to predict the scaling of intravalley velocity matrix elements, though it fails to reproduce the non-diagonal components $V_{0x}^i, V_{zy}^i$ and $V_{yz}^i$ of velocity matrix elements in inclined valleys.
In \KP\ the $V_{0x}^i$ component is magnified by $\times 10^3$ to make it visually appear in the plot.

\subsection{Dispersion}
\label{sec:dispersion}

\begin{figure}
  \includegraphics[width=0.9\linewidth]{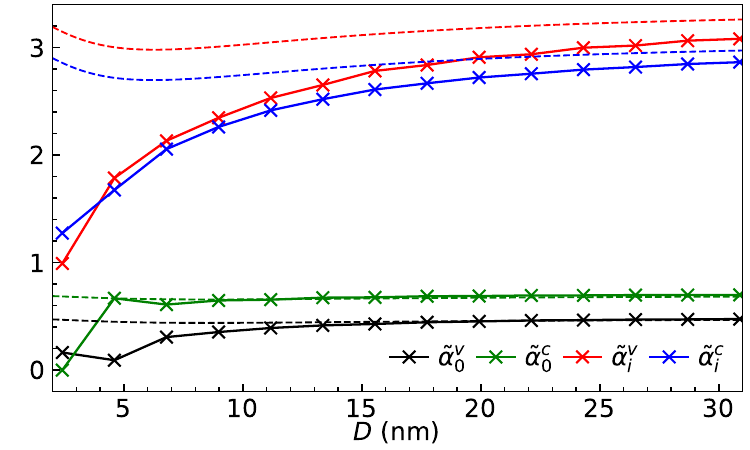}
  \caption{Scaling of the bare conduction (green and blue lines) and valence (black and red lines) band effective masses in longitudinal, $\tilde\alpha_{0}^{c(v)}$ and inclined, $\tilde\alpha_{i}^{c(v)}$, valleys as a function of the effective diameter in hexagonal PbS NWs.
  Dashed lines show the same values obtained in the framework of the anisotropic \KP\ model.
  }
  \label{fig:aaaa_PbS}
\end{figure}

Finally, we discuss the last term $\hat H^{c(v)}(k_z)$ in Eq.~\eqref{eq:Hb}.
This term is responsible for quadratic shift $\propto k_z^2$ of the electron (hole) energy levels.
It can be calculated in the framework of the \KP\ theory as the derivatives of matrix elements of the Hamiltonian at $k_z \ne 0$ between electron (hole) eigenstates.
This term does not depend on the valley mixing and is parametrized as follows
\begin{equation}\label{eq:Hzk}
  \hat H^{c(v)}(k_z) = \pm k_z^2 \diag(\tilde\alpha_0^{c(v)},\tilde\alpha_i^{c(v)},\tilde\alpha_i^{c(v)},\tilde\alpha_i^{c(v)})\otimes\mathbb 1,
\end{equation}
where $\tilde\alpha_0^{c(v)}$ describe the energy shift of electrons (holes) in longitudinal valley and $\tilde\alpha_i^{c(v)}$ in inclined valleys.
Parameters $\tilde\alpha_0^{c(v)}\ne\tilde\alpha_i^{c(v)}$ due to the different effective masses in longitudinal, $L_0$, and inclined, $L_i, i=1,2,3$, valleys along the NW axis.

The parameters $\tilde\alpha_{\mu}^b$ \eqref{eq:Hzk} are optimized to best fit the ETB NWs dispersion near the band extremum.
The other used parameters are the confinement energies and anisotropic splitting shown in Fig.~\ref{fig:EA_PbS_N}, the valley mixing constants shown in Fig.~\ref{fig:VS_PbS_N} and the velocity matrix elements $V_z$ shown in Fig.~\ref{fig:VME_PbS_N}.
The fitted values of $\alpha_{0}^{c(v)}$, $\alpha_{i}^{c(v)}$ are shown in Fig.~\ref{fig:aaaa_PbS} by solid lines with ``$\times$'' symbols indicating diameters at which the dispersion was fitted.
Calculations in the framework of the anisotropic \KP\ theory, see Appendix~\ref{app:kp}, are shown by dashed lines.
One can see that the bare effective mass parameters in longitudinal valley $\alpha_{0}^{c(v)}$ are almost insensitive to the change of the NW diameter.
Same parameters in inclined valleys are much larger $\alpha_{i}^{c(v)}$ and significantly change with increase of the NW diameter.

\begin{figure}
  \includegraphics[width=1\linewidth]{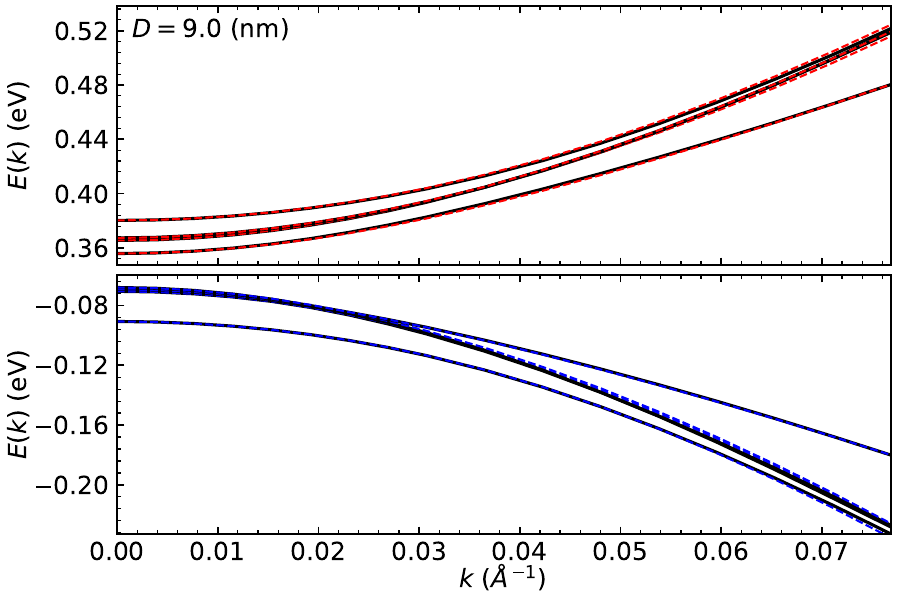}
  \caption{Dispersion in hexagonal PbS NW with size parameter $N=20$ and diameter $D\approx9$ nm in the vicinity of the band extremum at $k_z=0$. Results of tight-binding calculations are shown by solid black lines, calculations in the framework of the phenomenological model \eqref{eq:Hvc} are shown by red (blue) color for conduction (valence) band part of the dispersion.
  }
  \label{fig:HB_PbS_N20_disp}
\end{figure}

In Fig.~\ref{fig:HB_PbS_N20_disp} we show the electron and hole dispersion in the vicinity of the band extremum $k_z=0$ for the hexagonal PbS NW with the size parameter $N=20$ ($D\approx 9$ nm).
The ETB dispersion is shown by solid black lines and the dispersion of electron (hole) states calculated using the phenomenological model \eqref{eq:Hvc} is show by red (blue) dashed lines.

\section{Conclusions}

To conclude, using the valley untangling procedure we obtained parametrization for the valley mixing Hamiltonian in hexagonal and cylindrical PbS and PbSe NWs with $[111]$ growth direction.
The access to the valley states allow us to explicitly compute the valley mixing constants, components of the \gfs\ tensors and interband velocity matrix elements and compare them to the corresponding values calculated in the framework of the anisotropic \KP\ model.
We derived a simple phenomenological model which allows to accurately predict the fine structure of electron (hole) states not only at $k_z=0$, but in the vicinity of the band extremum.

We show that the constants of the valley mixing strongly depend on the microscopic structure of the NW surface and oscillate with the change of the NW diameter.
In nanowires without center of inversion additional valley mixing constants are responsible for the lineark in $k$ spin splittings of the electron and hole subbands.
The other physical properties, such as velocity matrix elements, \gfs\ tensors or anisotropic splittings are smooth functions of the NW diameter and can be calculated analytically with fair accuracy.
The data for cylindrical PbS and cylindrical and hexagonal PbSe NWs are given in supplemental information in Appendix~\ref{app:SI}.

\section*{Acknowledgments}
The work of IDA was supported by Russian Science Foundation under grant no. 22-72-00121.
IDA also thanks the Foundation for Advancement of Theoretical Physics and Mathematics ``BASIS''.

\begin{appendix}

\section{Symmetrization matrices}
\label{app:symm}

In this section we define the electron (hole) valley states in the $[111]$-grown PbX nanowires with $D_{3d}$ symmetry and derive and their transformation and symmetrization matrices.
The matrices from Ref.~\cite{Avdeev2017} are not reused since they were derived in the assumption of independent valley and spin degrees of freedom.
This assumption is a valid option for symmetry analysis, however it unnecessary complicates the matrix elements of operators since spinors in inclined valleys become misaligned with the corresponding valley axes.
Here we follow Refs.~\cite{Avdeev2020,Avdeev2023g} and relate the states in different valleys by symmetry operations of the NW point group $D_{3d}$.
The parity of the states (the band index) is omitted throughout the section since transformation $T(c)=\pm\mathbb1$ under the inversion $c\in D_{3d}$ is trivial and technically it is enough to consider only $D_3$ subgroup of the $D_{3d}$. 
Details of the point symmetry of the NWs are given in Sec.~\ref{sec:NW}.

The valley pseudospinors are defined as follows.
The basis functions in the longitudinal valley $L_0$,
\begin{equation}
  \ebasis_0 = (\ket{0\up},\ket{0\down}),
\end{equation}
are assumed to be independent from the bases in inclined valleys and transform via $g\ebasis_0=\ebasis_0T(g)$ as spinors
\begin{equation}
  \label{eq:T0_ab}
  T_{0}(a) = 
  \begin{pmatrix}
    \omega^* & 0 \\
    0 & \omega \\
  \end{pmatrix}
  ,~
  T_{0}(b) = -\rmi \sigma_x
  ,~
  \omega = \rme^{\frac{\rmi\pi}{3}},
\end{equation}
under the generators $g=a,b$ of the NW point group $D_{3d}$.
Indeed, there is no symmetry operation $g\in D_{3d}$ which transforms the longitudinal valley $L_0$ into one of inclined valleys $L_i, i=1,2,3$, see Eq.~\eqref{eq:LL}.
The basis functions in the inclined valley $L_1$,
\begin{equation}
  \ebasis_1 = (\ket{1\up},\ket{1\down}),
\end{equation}
are also assumed to transform as spinors under the generator $b=C_{2x_0}\in D_{3d}$.
The bases in inclined valleys $L_2$ and $L_3$ are defined via $C_{3z_0}$ rotation \eqref{eq:cf0} of the $\ebasis_1$
\begin{equation}\label{eq:E23E1}
  \ebasis_2 = C_{3z_0}\ebasis_1,\quad
  \ebasis_3 = C_{3z_0}^2\ebasis_1.
\end{equation}
This yields the following transformation matrices
\begin{equation}
  \label{eq:Ti_ab}
  T_{i}(a) = 
  \begin{pmatrix}
    \mathbb 0 & \mathbb 0 &- \mathbb 1 \\
    \mathbb 1 & \mathbb 0 &  \mathbb 0 \\
    \mathbb 0 & \mathbb 1 &  \mathbb 0 \\
  \end{pmatrix}
  ,\quad
  T_{i}(b) = 
  \begin{pmatrix}
    -\rmi& 0 & 0 \\
    0 & 0 &\rmi \\
    0 &\rmi& 0 \\
  \end{pmatrix}  
  \otimes \sigma_x.
\end{equation}
To derive these transformation matrices let $g_{\mu\to \nu} \in D_{3d}$ transform the valley $L_{\mu}\to L_{\nu}$ ($C_{3z_0}$ transforms $L_1\to L_2 \to L_3\to L_1$, see Fig.~\ref{fig:NW_N6}).
Let also $g_{\lambda\to\mu}\ebasis_{\lambda}=\ebasis_{\mu}$ and $g_{\lambda\to\nu}\ebasis_{\lambda}=\ebasis_{\nu}$.
For inclined valleys $\lambda=1$.
Then for $g_{\mu \to \nu}: \ebasis_{\mu} \to \ebasis_{\nu}$ one has
\begin{equation}\label{eq:T_blk}
  g_{\mu\to \nu} \left(g_{\lambda\to \mu} \ebasis_{\lambda}\right) = \left(g_{\lambda\to \nu} \ebasis_{\lambda} \right) T_{\mu\to \nu}(g_{\mu\to \nu}),
\end{equation}
where $T_{\mu\to \nu}(g_{\mu\to \nu})$ are the blocks of the transformation matrices \eqref{eq:Ti_ab}.
Therefore in the coordinates frame where $\ebasis_{\lambda}$ transforms as spinors the block reads as
\begin{equation}
  \label{eq:T_block}
  T_{\mu\to\nu}(g_{\mu\to\nu}) = D_{\frac12}^{\dag}(g_{\lambda\to \nu})D_{\frac12}(g_{\mu\to \nu})D_{\frac12}(g_{\lambda\to \mu}),
\end{equation}
where $D_{\frac12}(\bm n,\omega)=\cos(\frac{\omega}2)-\rmi\bm n\bm \sigma\sin(\frac{\omega}2)$ is the spin rotation matrix.
As discussed in Sec.~\ref{sec:NW}, for $\lambda=1$ the NW coordinates frame \eqref{eq:cf0} is used.

The matrices \eqref{eq:T0_ab} are the standard matrices \cite{Koster} of the $\Gamma_4$ irreducible representation of the $D_3\subset D_{3d}$.
The matrices \eqref{eq:Ti_ab} are the matrices of $2\Gamma_4\oplus\Gamma_5\oplus\Gamma_6$ reducible representation of the $D_3\subset D_{3d}$.
They can be brought to the standard block-diagonal form using either $S_i$ or $U_iS_i$ symmetrization matrices.
The $S_i$ matrix is
\begin{equation}
  \label{eq:Si}
  S_{i} = 
  \begin{pmatrix}
    \frac{{\omega^*}^3}{\sqrt3} & 0 & 0 & \frac{\rmi{\omega}^3}{\sqrt3} & \frac{\rmi}{\sqrt6} & \frac{\rmi}{\sqrt6} \\ 
    0 & \frac{{\omega^*}^3}{\sqrt3} & \frac{\rmi{\omega^*}^3}{\sqrt3} & 0 & \frac{\rmi}{\sqrt6} &-\frac{\rmi}{\sqrt6} \\
    \frac{{\omega^*}^2}{\sqrt3} & 0 & 0 & \frac{\rmi{\omega}^2}{\sqrt3} &-\frac{\rmi}{\sqrt6} &-\frac{\rmi}{\sqrt6} \\
    0 & \frac{{\omega^*}^2}{\sqrt3}  & \frac{\rmi{\omega^*}^2}{\sqrt3} & 0 &-\frac{\rmi}{\sqrt6} & \frac{\rmi}{\sqrt6} \\
    \frac{{\omega^*}  }{\sqrt3} & 0 & 0 & \frac{\rmi{\omega}  }{\sqrt3}  & \frac{\rmi}{\sqrt6} & \frac{\rmi}{\sqrt6} \\
    0 & \frac{{\omega^*}}{\sqrt3} & \frac{\rmi{\omega^*}}{\sqrt3} & 0 & \frac{\rmi}{\sqrt6} &-\frac{\rmi}{\sqrt6} \\
  \end{pmatrix}
  ,~
  \omega=\rme^{\frac{\rmi\pi}{3}},
\end{equation}
and the matrix $U_i$,
\begin{equation}\label{eq:Ui}
  U_i=\rmi \diag(\sigma_x,\sigma_x,\sigma_x),
\end{equation}
commutes with both $T_i(a)$ and $T_i(b)$.
The phase $\rmi$ is chosen to satisfy the time inversion symmetry.

Since the states in longitudinal and inclined valleys are chosen as independent, then their relative sign gives another degree of freedom.
When both band are considered, then the overall sign of conduction (valence) band function should be also taken into account.
Therefore the general form of the symmetrization matrix becomes
\begin{equation}
  \label{eq:S}
  S = 
  \pm
  \begin{pmatrix}
    \mathbb 1 & \mathbb 0 \\
    \mathbb 0 & \pm[U_i]S_i
  \end{pmatrix}.
\end{equation}

\section{Anisotropic effective mass theory}
\label{app:kp}

\subsection{Eigenstates}

We propose three index (fourth is $k_z$, optional) notation for the nanowire bispinor eigenstates \cite{Avdeev2017} in the NW-related cylindrical coordinates frame $(\rho,\varphi,z)$
\begin{equation}\label{eq:MNpk}
  \ket{MNp[k_z]} = \rme^{\rmi k_zz} \left[ u_{MN}(\rho) \Omega_{Mp}^{-}(\varphi) \ket{-} + \rmi v(\rho) \Omega_{Mp}^{+} \ket{+}\right],
\end{equation}
where $\ket{\mp}$ is the conduction (valence) odd (even) Bloch function, $u(v)$ are conduction (valence) band radial envelope and $\Omega_{Mp}^s(\varphi), s=\mp1$ is the angular part of the wave function, an eigenfunction of $\hat J_z = \hat s_z + \hat L_z$.
The quantum numbers are: $M=\pm\frac12,\pm\frac32,\ldots$ is the half-integer projection of total angular momentum, $N$ enumerates roots of dispersion equation for certain $M$ and $p=\pm1$ or $\up(\down)$ is the spin-like index.

The functions $\Omega^{\pm}\equiv\Omega^s$ are
\begin{equation}\label{eq:Omega}
  \Omega_{Mp}^s(\varphi) = \frac{1}{2\sqrt{2\pi}}\rme^{\rmi p(M+\frac{s}{2})\varphi} \begin{pmatrix} 1-sp \\ 1+sp \end{pmatrix},~s=\pm1.
\end{equation}
The orthogonality properties of $\Omega$ are given by
\begin{equation}
  \brakt{s'M'p'}{sMp} = \delta_{s'p',sp} \delta_{s'M',sM}.
\end{equation}
The functions \eqref{eq:Omega} allow to easily calculate the matrix elements of different $\hat s$, $\hat L_z$ and $\hat s \cdot \hat L_z$ operators.

The radial envelopes are
\begin{subequations}
  \begin{equation}
    u_{MN}(\rho) = N_{MN}[J_{M-\frac12}(k\rho)+c I_{M-\frac12}(\varkappa\rho)],
  \end{equation}
  \begin{equation}
    v_{MN}(\rho) = N_{MN}[g J_{M+\frac12}(k\rho)+c G I_{M+\frac12}(\varkappa\rho)],
  \end{equation}
\end{subequations}
where
\begin{subequations}
\begin{equation}
  c = -\frac{J_{M-\frac12}(kR)}{I_{M-\frac12}(\varkappa R)},
\end{equation}
\begin{equation}
  g = \frac{Pk}{\alpha_v k^2+E+\frac{E_g}2},
\end{equation}
\begin{equation}
  G = \frac{P\varkappa}{\alpha_v \varkappa^2-E-\frac{E_g}2},
\end{equation}
\begin{equation}
  k = \sqrt{\Xi+\Lambda},\quad \varkappa = \sqrt{\Xi-\Lambda},
\end{equation}
\begin{equation}
  \Xi = \sqrt{\Lambda^2+\frac{4E^2-E_g^2}{4\alpha_v\alpha_c}},
\end{equation}
and
\begin{equation}
  \Lambda = \frac{E(\alpha_v-\alpha_c)-P^2-(\alpha_v+\alpha_c)\frac{E_g}2}{2\alpha_v\alpha_c}.
\end{equation}
\end{subequations}
Derivatives of the radial functions $u,v$ are calculated using the following relations for derivatives of Bessel and modified Bessel functions
\begin{equation}
  J'_{\nu} = \frac12(J_{\nu-1}-J_{\nu+1})
  ,\quad
  I'_{\nu} = \frac12(I_{\nu-1}+I_{\nu+1}).
\end{equation}

\begin{widetext}
The 1st order derivatives of the products of $u,v$ and $\Omega$ are 
\begin{subequations}\label{eq:d1hO}
  \begin{equation}
      \nabla_0 h \Omega_{Mp}^s = \rmi k_z h \Omega_{Mp}^s,
  \end{equation}
  \begin{equation}
    \nabla_{\pm} h \Omega_{Mp}^s = \frac{1}{\sqrt2}\left(\mp h'+\frac{pM_s h}{\rho}\right)\left[\Omega_{M\pm p,p}^s=\rme^{\pm\rmi\varphi}\Omega_{M,p}^s\right].
  \end{equation}
\end{subequations}
The 2nd order derivatives are
\begin{subequations}\label{eq:d2hO}
  \begin{equation}
    \nabla_+\nabla_+ h \Omega_{Mp}^s = \frac12\left(+h''-\frac{(1+2pM_s) h'}{\rho}+\frac{pM_s(pM_s+2)h}{\rho^2}\right)\left[\Omega_{M+2p,p}^s=\rme^{+2\rmi\varphi}\Omega_{M,p}^s\right],
  \end{equation}
  \begin{equation}
    \nabla_-\nabla_+ h \Omega_{Mp}^s = \frac12\left(-h''-\frac{h'}{\rho}+\frac{M_s^2 h}{\rho^2}\right)\Omega_{M,p}^s,
  \end{equation}
  \begin{equation}
    \nabla_-\nabla_- h \Omega_{Mp}^s = \frac12\left(+h''-\frac{(1-2pM_s) h'}{\rho}+\frac{pM_s(pM_s-2)h}{\rho^2}\right)\left[\Omega_{M-2p,p}^s=\rme^{-2\rmi\varphi}\Omega_{M,p}^s\right].
  \end{equation}
\end{subequations}
Here $M_s = M+\frac{s}2$ and $h\equiv h(\rho)$.
\end{widetext}

\subsection{Generalized form of the Hamiltonian}
Generalized form of the \KP\ Hamiltonian reads as
\begin{equation}
  \label{eq:H_an_rot}
  \hat H =
  \begin{pmatrix}
    \frac{E_g}2 - \alpha_c^{ij} \nabla_i\nabla_j & -\frac{\rmi\hbar}{m_0} P_{ij} \sigma_i\nabla_j \\
    -\frac{\rmi\hbar}{m_0} P_{ij} \sigma_i\nabla_j & -\frac{E_g}2 + \alpha_v^{ij}\nabla_i\nabla_j
  \end{pmatrix}.
\end{equation}
Here $\alpha^{ij}=\alpha^{ji}$ since $\nabla_i\nabla_j=\nabla_j\nabla_i$, while in general $P_{ij}\ne P_{ji}$ since $\sigma_i\nabla_j \ne \sigma_j\nabla_i$.
The key to compute the matrix elements of this operator is to convert cartesian derivatives and Pauli matrices to cyclic ones.
\begin{widetext}
In cyclic coordinates the terms with $\alpha$ tensors and derivatives read as:
\begin{multline}
  \alpha^{ij}\nabla_i\nabla_j = 
  \alpha^{xx}\nabla_x\nabla_x+
  \alpha^{yy}\nabla_y\nabla_y+
  \alpha^{zz}\nabla_z\nabla_z+
  2\left(
  \alpha^{xy}\nabla_x\nabla_y+
  \alpha^{xz}\nabla_x\nabla_z+
  \alpha^{yz}\nabla_y\nabla_z
  \right)
  = \\
  \left(\frac{\alpha^{xx}-\alpha^{yy}}2-\rmi\alpha^{xy}\right)\nabla_+\nabla_+
  +\sqrt2\left(-\alpha^{xz}+\rmi\alpha^{yz}\right)\nabla_0\nabla_+
  +\alpha^{zz}\nabla_0\nabla_0
  + \\
  -(\alpha^{xx}+\alpha^{yy})\nabla_-\nabla_+
  +
  \sqrt2\left(\alpha^{xz}+\rmi\alpha^{yz}\right)\nabla_0\nabla_-
  +
  \left(\frac{\alpha^{xx}-\alpha^{yy}}2+\rmi\alpha^{xy}\right)\nabla_-\nabla_-\,.
\end{multline}
Similar expression are obtained for $P_{ij}\sigma_i\nabla_j$:
\begin{multline}
  \sum_{i,j=x,y,z} P_{ij}\sigma_i\nabla_j =
  \frac{P_{xx}-P_{yy}-\rmi(P_{xy}+P_{yx})}2
  \sigma_+\nabla_+
  -\frac{P_{xz}-\rmi P_{yz}}{\sqrt2}\sigma_+\nabla_0
  -\frac{P_{zx}-\rmi P_{zy}}{\sqrt2}\sigma_0\nabla_+
  + \\ +
  P_{zz}\sigma_0\nabla_0
  -\frac{P_{xx}+P_{yy}+\rmi(P_{xy}-P_{yx})}2\sigma_+\nabla_-
  -\frac{P_{xx}+P_{yy}-\rmi(P_{xy}-P_{yx})}2\sigma_-\nabla_+
  + \\
  +\frac{P_{xz}+\rmi P_{yz}}{\sqrt2}\sigma_-\nabla_0
  +\frac{P_{zx}+\rmi P_{zy}}{\sqrt2}\sigma_0\nabla_-
  +\frac{P_{xx}-P_{yy}+\rmi(P_{xy}+P_{yx})}2\sigma_-\nabla_-
  \,.
\end{multline}
For NWs $\nabla_z \to \rmi k_z$ and $k_z = 0$ at the band extremum.
\end{widetext}

\subsection{Generalized form of velocity operator}
\label{subsec:velocity}

The generalized form of the velocity operator is
\begin{equation}
  \hat{v}_k = \frac{\rmi}{\hbar} \left[\hat H, \hat{r}_k \right] =
  \begin{pmatrix}
    -\frac{2\rmi}{\hbar}\alpha_c^{jk}\nabla_j &
    \frac{P_{jk}}{m_0}\sigma_j \\
    \frac{P_{jk}}{m_0}\sigma_j &
    \frac{2\rmi}{\hbar}\alpha_v^{jk}\nabla_j
  \end{pmatrix}.
\end{equation}
Its matrix elements between the basis wave functions $\ket{MNp}$, Eq.~\eqref{eq:MNpk}, can be easily calculated using the explicit form of 1st order derivatives \eqref{eq:d1hO}.

\subsection{Interaction with magnetic field}
\label{app:B}

Interaction with magnetic field is calculated as described in supplemental information for Ref.~\cite{Avdeev2023g} with the vector potential given by Eq.~\eqref{eq:A}.
This gauge reduces the problem of interaction with magnetic field to two dimensions.
The vector potential \eqref{eq:A} takes the following form in cyclic coordinates
\begin{equation}
  A_{\pm} = -\rmi\frac{B_z}{2} \frac{x\pm\rmi y}{\sqrt{2}},\quad
  A_0 = yB_x-xB_y.
\end{equation}
It is convenient to represent the vector potential as
\begin{equation}
  \label{eq:tA} 
  A_{\pm} = \tilde A_0 \rho \rme^{\pm\rmi\varphi},\quad
  A_0 = \tilde A_- \rho \rme^{+\rmi\varphi} + \tilde A_+ \rho \rme^{-\rmi\varphi},
\end{equation}
where
\begin{equation}
  \label{eq:tAc} 
  \tilde A_{\pm} = \pm\frac{\rmi}2\left([B_x\pm\rmi B_y]\right),\quad
  \tilde A_0 = -\frac{\rmi}{2\sqrt2} B_z.
\end{equation}
Finally, to avoid non-Hermitian terms in the Hamiltonian one should symmetrize the 2nd derivatives 
\[
  \nabla_s\nabla_{s'}=\frac12(\nabla_s\nabla_{s'}+\nabla_{s'}\nabla_s)
  ,\quad
  s,s'=0,\pm1
\]
to obtain the following linear-in-$B$ corrections $\propto \alpha$ to the Hamiltonian
\begin{equation}
  \mp \alpha_{c(v)}^{ss'}
  \frac{\rmi e}{2\hbar c}(A_s\nabla_{s'}+\nabla_{s'}A_s+A_{s'}\nabla_s+\nabla_sA_{s'}).
\end{equation}
This expression can be further simplified using the commutations relations for the vector potential \eqref{eq:A}
\begin{equation}
  \left[\nabla_{\pm},A_{\mp}\right]=\pm\frac{\rmi B_z}{2}
  ,\quad
  \left[\nabla_{\pm},A_0\right]=-\rmi \frac{B_x{\pm}\rmi B_y}{2}.
\end{equation}
All others commutators are zero.
Linear-in-$B$ terms $\propto P$ do not require symmetrization and are simpler to compute since they do not contain derivatives.

\subsection{Parameters for the effective mass model}
\label{subsec:kp_params}

Material parameters \cite{Avdeev2023g} used for the \KP\ calculations throughout the paper are listed in Table~\ref{tb:kp_params} in atomic units $\hbar=m_0=|e|=1$.
For convenience the low temperature band gap of PbS and PbSe are also given in eV.
The bare electron (hole) $g$-factors (the remote band contributions to the Zeeman splitting) are related to the bulk ones as follows
\begin{subequations}
  \begin{equation}
    g_{0t}^c = g_{t,\text{bulk}}^c - \frac{4P_tP_l}{E_gm_0}
    ,\quad
    g_{0l}^c = g_{l,\text{bulk}}^c - \frac{4P_t^2}{E_gm_0},
  \end{equation}
  \begin{equation}
    g_{0t}^v = g_{t,\text{bulk}}^v + \frac{4P_tP_l}{E_gm_0}
    ,\quad
    g_{0l}^v = g_{l,\text{bulk}}^v + \frac{4P_t^2}{E_gm_0},
  \end{equation}
\end{subequations}
see Ref.~\cite{Avdeev2023g} for details.

\begin{table}
\caption{Material parameters for anisotropic \KP\ model. All parameters are given in atomic units $\hbar=m_0=|e|=1$. $E_g$ is also given in eV.}
\label{tb:kp_params}
\begin{ruledtabular}
\begin{tabular}{c|cc}
  & PbS & PbSe \\
  \hline
  $\alpha_v^t$          & $ 3.71261$ & $3.61848$ \\
  $\alpha_v^l$          & $ 0.48082$ & $0.78444$ \\
  $\alpha_c^t$          & $ 3.35927$ & $3.00629$ \\
  $\alpha_c^l$          & $ 0.69673$ & $0.94606$ \\
  \hline
  $P_t$                 & $ 0.17556$ & $0.26986$ \\
  $P_l$                 & $ 0.21101$ & $0.22430$ \\
  \hline
  $E_g$                 & $ 0.01080$ & $0.00782$ \\
  $E_g$ (eV)            & $ 0.29397$ & $0.21288$ \\
  \hline
  $g_{v,\text{bulk}}^t$ & $-9.62387$ & $-24.19664$ \\
  $g_{v,\text{bulk}}^l$ & $-9.99538$ & $-31.45314$ \\
  $g_{c,\text{bulk}}^t$ & $10.13564$ & $ 25.99249$ \\
  $g_{c,\text{bulk}}^l$ & $11.05293$ & $ 31.26533$ \\
\end{tabular}%
\end{ruledtabular}%
\end{table}

\end{appendix}

\bibliography{nw} 



\clearpage
\newpage
\onecolumngrid

\begin{center}
  \textsf{\textbf{\Large Supplementary Information }} \\
\end{center}
\setcounter{equation}{0}
\setcounter{figure}{0}
\setcounter{table}{0}
\setcounter{page}{1}
\setcounter{section}{0}
\makeatletter
\renewcommand{\thepage}{S\arabic{page}}
\renewcommand{\theequation}{S\arabic{equation}}
\renewcommand{\thefigure}{S\arabic{figure}}
\renewcommand{\thetable}{S\arabic{table}}
\renewcommand{\thesection}{S\arabic{section}}


\section{Results for \NoCaseChange{PbSe} and cylindrical NWs}
\label{app:SI}

In this section we show remaining results of our calculations for hexagonal PbSe NWs and for cylindrical PbS and PbSe NWs with $D_{3d}$ symmetry.
Anisotropic confinement energies, velocity matrix elements, components of the \gfs\ tensors and remote band contribution to the electron (hole) effective masses are shown only for hexagonal NWs since in cylindrical ones they are almost the same.

\begin{figure}[h]
  \includegraphics[width=0.8\linewidth]{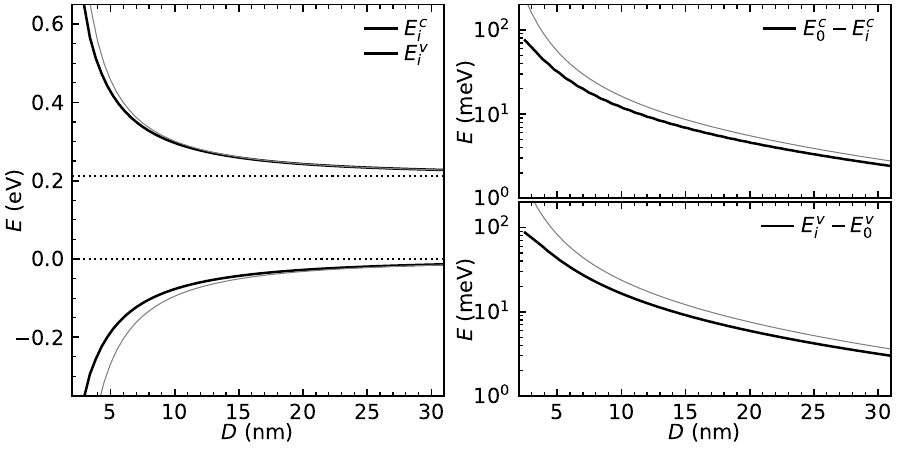}
  \caption{Same as in Fig.~\ref{fig:EA_PbS_N}, but for hexagonal PbSe NWs.}
  \label{fig:EA_PbSe_N}
\end{figure}

\begin{figure}[h]
  \includegraphics[width=0.8\linewidth]{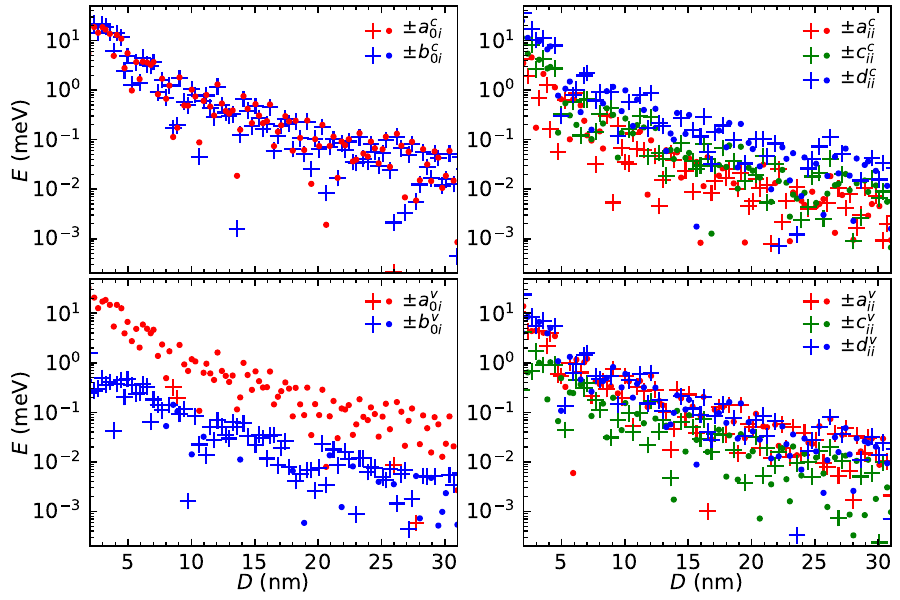}
  \caption{Same as in Fig.~\ref{fig:VS_PbS_N}, but for cylindrical PbS NWs.}
  \label{fig:VS_PbS_R}
\end{figure}
\begin{figure}[h]
  \includegraphics[width=0.8\linewidth]{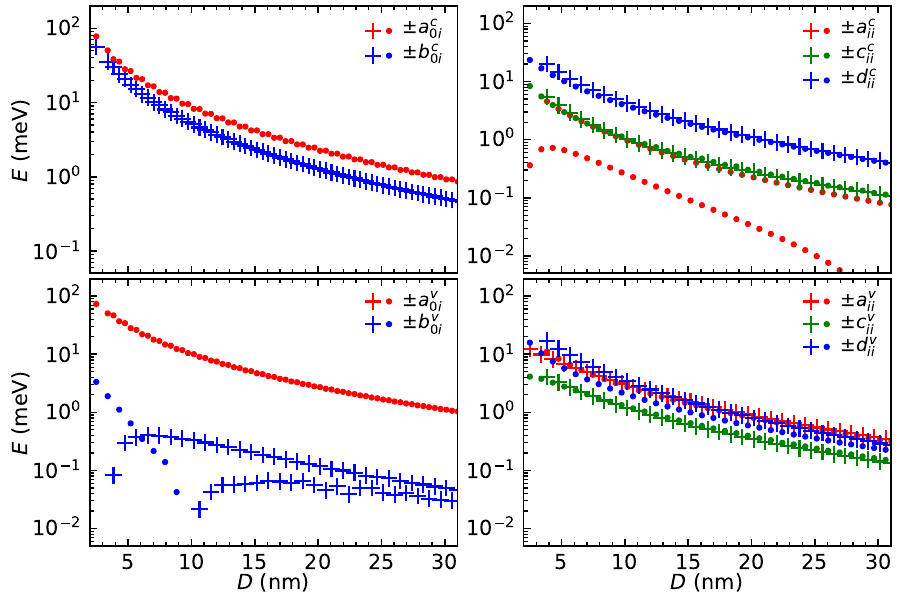}
  \caption{Same as in Fig.~\ref{fig:VS_PbS_N}, but for hexagonal PbSe NWs.}
  \label{fig:VS_PbSe_N}
\end{figure}
\begin{figure}[h]
  \includegraphics[width=0.8\linewidth]{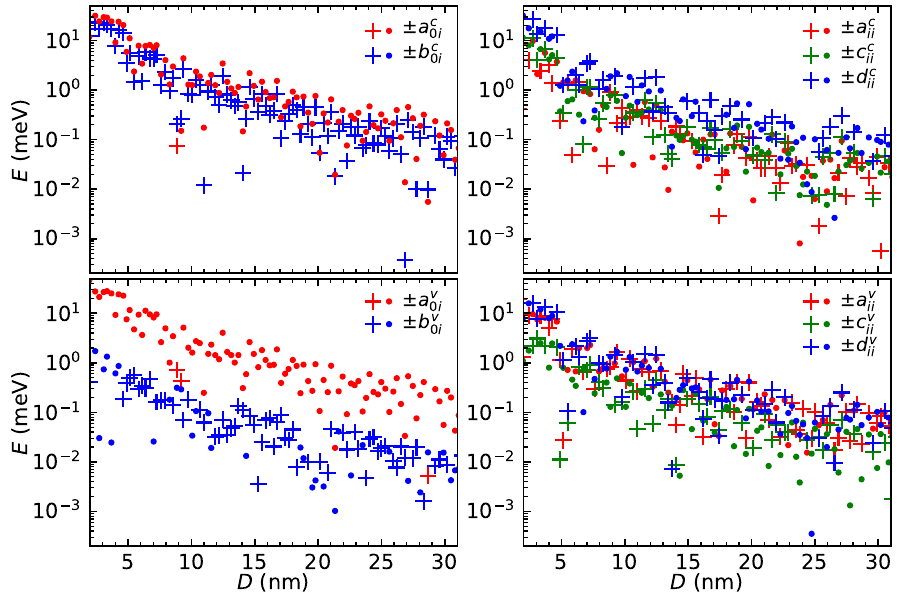}
  \caption{Same as in Fig.~\ref{fig:VS_PbS_N}, but for cylindrical PbSe NWs.}
  \label{fig:VS_PbSe_R}
\end{figure}

\begin{figure}[h]
  \includegraphics[width=0.8\linewidth]{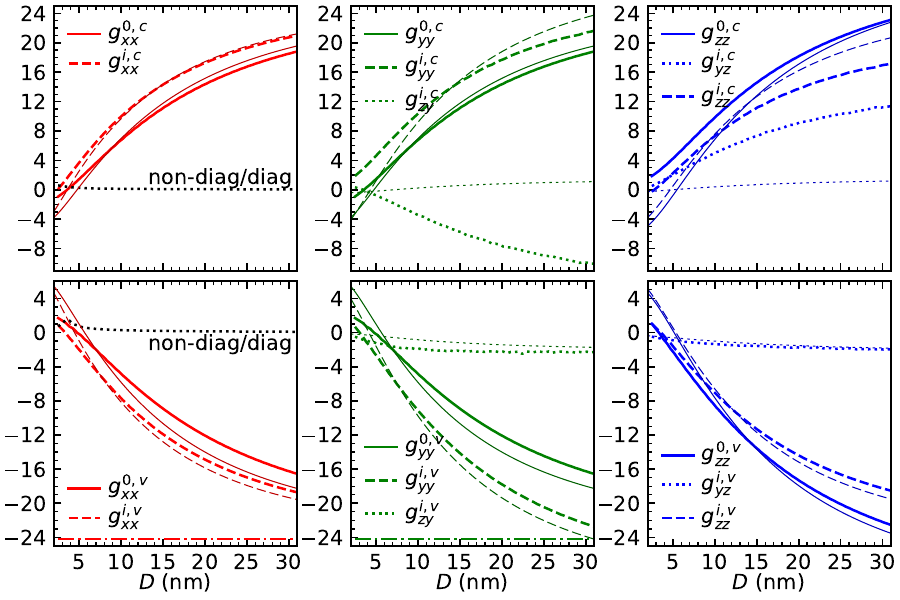}
  \caption{Same as in Fig.~\ref{fig:HB_PbS_N}, but for hexagonal PbSe NWs.}
  \label{fig:HB_PbSe_N}
\end{figure}

\begin{figure}[h]
  \includegraphics[width=0.8\linewidth]{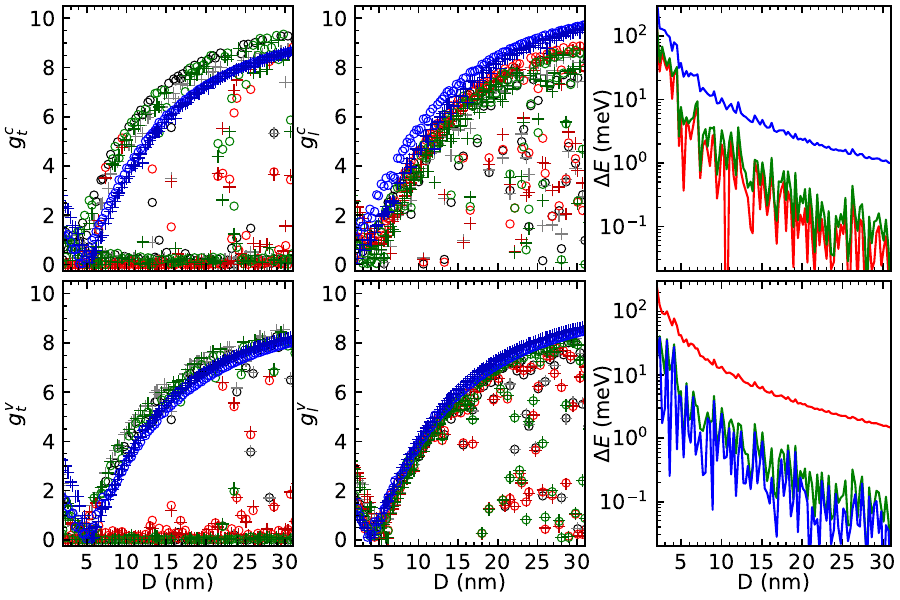}
  \caption{Same as in Figs.~\ref{fig:gg_HB_even},~\ref{fig:gg_HB_odd}, but for cylindrical PbS NWs (without odd-even separation).
  The tight-binding calculations of the electron (hole) \gfs\ are shown by ``$o$'' symbols, results of the calculation within the framework of the phenomenological model \eqref{eq:Hvc} with the intravalley tensors of \gfs\ calculated in \KP, see Appendix \ref{app:B}, are shown by ``$+$'' symbols.}
  \label{fig:gg_HB_PbS_R}
\end{figure}
\begin{figure}[h]
  \includegraphics[width=0.8\linewidth]{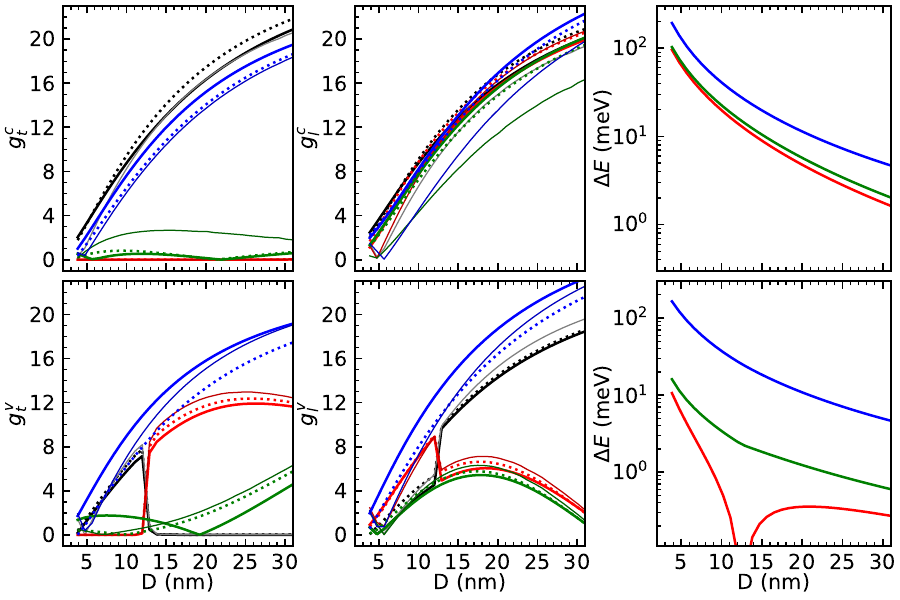}
  \caption{Same as in Fig.~\ref{fig:gg_HB_even}, but for hexagonal PbSe NWs.}
  \label{fig:gg_HB_PbSe_even}
\end{figure}
\begin{figure}[h]
  \includegraphics[width=0.8\linewidth]{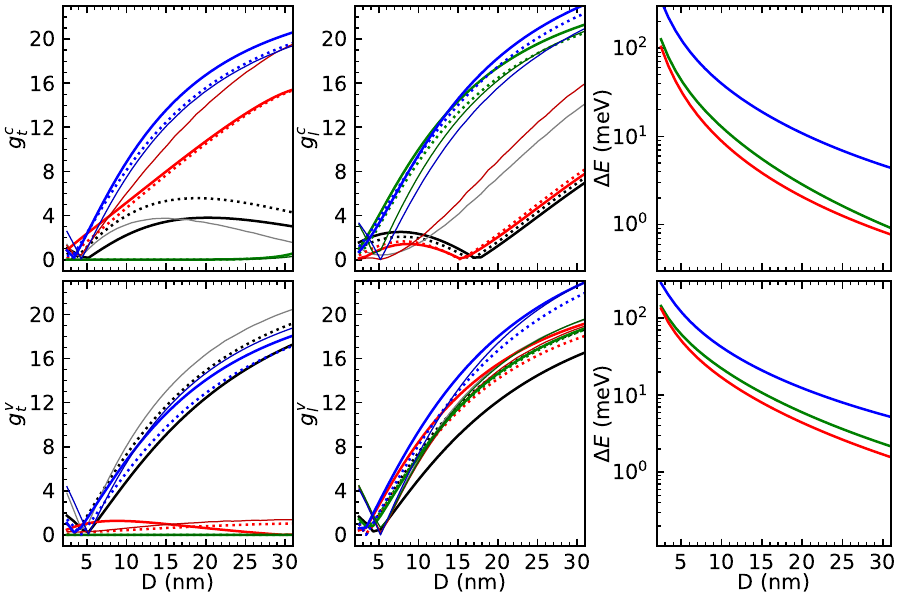}
  \caption{Same as in Fig.~\ref{fig:gg_HB_odd}, but for hexagonal PbSe NWs.}
  \label{fig:gg_HB_PbSe_odd}
\end{figure}
\begin{figure}[h]
  \includegraphics[width=0.8\linewidth]{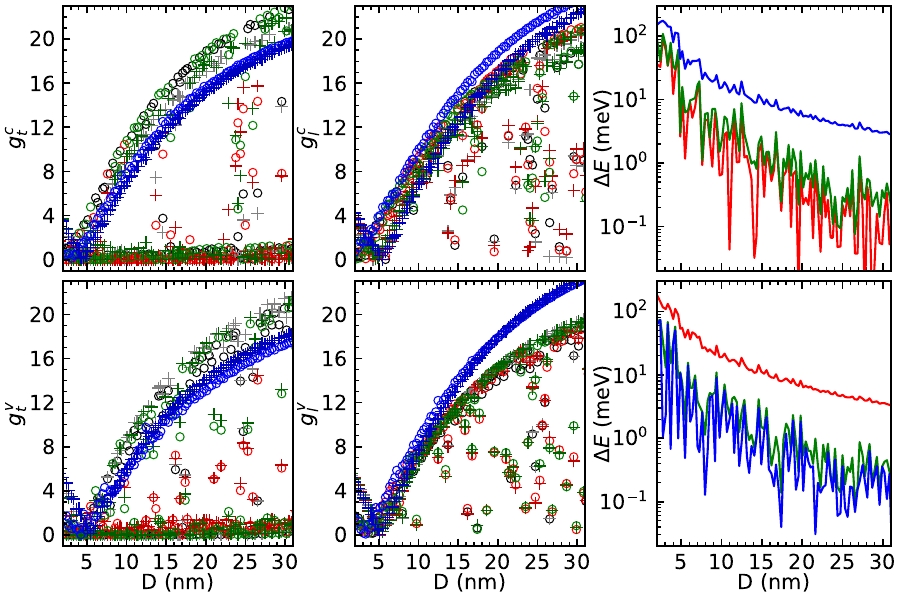}
  \caption{Same as in Fig.~\ref{fig:gg_HB_PbS_R}, but for cylindrical PbSe NWs.}
  \label{fig:gg_HB_PbSe_R}
\end{figure}

\begin{figure}[h]
  \includegraphics[width=0.8\linewidth]{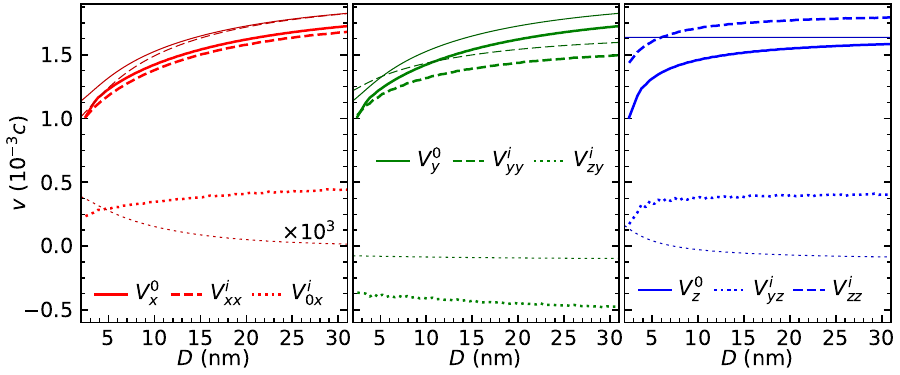}
  \caption{Same as in Fig.~\ref{fig:VME_PbS_N}, but for hexagonal PbSe NWs.}
  \label{fig:VME_PbSe_N}
\end{figure}

\begin{figure}[h]
  \includegraphics[width=0.6\linewidth]{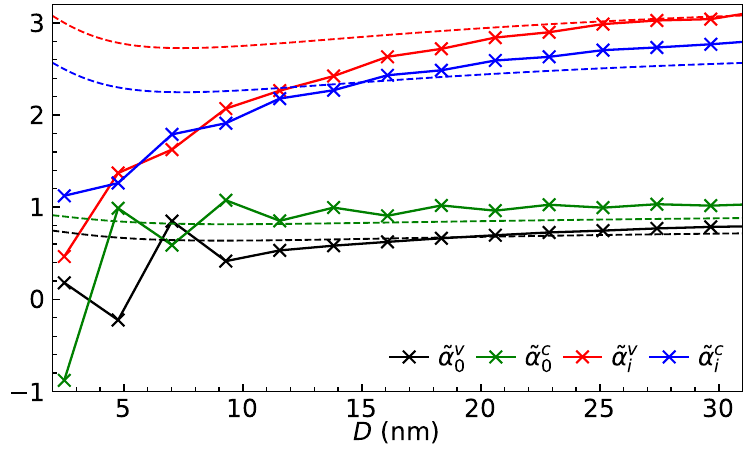}
  \caption{Same as in Fig.~\ref{fig:aaaa_PbS}, but for hexagonal PbSe NWs.}
  \label{fig:aaaa_PbSe}
\end{figure}

\end{document}